\documentclass[12pt,letterpaper]{article}

\usepackage{algorithm,algorithmic}
\usepackage{amsmath,amsfonts,amsthm,xcolor}
\usepackage[numbers]{natbib}
\usepackage{graphicx}
\usepackage{float, dsfont}
\usepackage{subfig}
\usepackage{verbatim}
\usepackage{hyperref}

\newcommand{\blind}{1}

\newcommand{\tcb}{}

\newcommand{\argmin}{\mathop{\mathrm{argmin}}}

\newcommand{\bb}{\mathbf}
\newcommand{\RR}{\mathbb{R}}

\begin{document}

\def\spacingset#1{\renewcommand{\baselinestretch}%
{#1}\small\normalsize} \spacingset{1}

%%%%%%%%%%%%%%%%%%%%%%%%%%%%%%%%%%%%%%%%%%%%%%%%%%%%%%%%%%%%%%%%%%%%%%%%%%%%%%

\if1\blind
{
  \title{Gradient boosting for extreme quantile regression}
\author{
Jasper Velthoen \footnote{Department of Applied Mathematics, Delft University of Technology, Mekelweg 4 2628 CD Delft. E-mail: \url{j.j.velthoen@tudelft.nl}}
\and Cl\'ement Dombry \footnote{ Universit{\'e} Bourgogne Franche-Comt{\'e}, Laboratoire de Math{\'e}matiques de Besan\c{c}on,  CNRS UMR 6623, F-25000 Besan\c{c}on, France. E-mail: \url{clement.dombry@univ-fcomte.fr}} 
\and Juan-Juan Cai \footnote{ Department of Econometrics and Data Science, Vrije Universiteit Amsterdam,
De Boelelaan 1105, 1081HV, Amsterdam, the Netherlands. E-mail: \url{j.cai@vu.nl}}
\and Sebastian Engelke\footnote{Research Center for Statistics, University of Geneva, Boulevard du Pont d’Arve 40, 1205 Geneva, Switzerland.  E-mail: \url{sebastian.engelke@unige.ch}}
}  
  \maketitle
} \fi

\if0\blind
{
  \bigskip
  \bigskip
  \bigskip
  \begin{center}
    {\LARGE\bf Gradient boosting for extreme quantile regression}
\end{center}
  \medskip
} \fi

%\bigskip
\begin{abstract}
  Extreme quantile regression provides estimates of conditional quantiles outside the range of the data. \tcb{Classical quantile regression performs poorly in such cases since data in the tail region are too scarce. Extreme value theory is used for extrapolation beyond the range of observed values  and estimation of conditional extreme quantiles. Based on the peaks-over-threshold approach, the conditional distribution above a high threshold is approximated by a generalized Pareto distribution with covariate dependent parameters.}  We propose a gradient boosting procedure to estimate a conditional generalized Pareto distribution by minimizing its deviance. Cross-validation is used for the choice of tuning parameters such as the number of trees and the tree depths. We discuss diagnostic plots such as variable importance and partial dependence plots, which help to interpret the fitted models. In simulation studies we show that our gradient boosting procedure outperforms classical methods from quantile regression and extreme value theory, especially for high-dimensional predictor spaces and complex parameter response surfaces. An application to statistical post-processing of weather forecasts with precipitation data in the Netherlands is proposed. 
\end{abstract}
\noindent%

\spacingset{1.5} % DON'T change the spacing!

\section{Introduction}
In a regression setup the distribution of a quantitative response $Y$ depends on a set of covariates (or predictors) $\bb X \in \mathbb R^d$. These predictors are typically easily available and can be used to predict conditional properties of the response variable $Y$. Machine learning offers a continuously growing set of tools to perform prediction tasks based on a sample $(\bb X_1,Y_1),\dots ,(\bb X_n,Y_n)$ of independent copies of a random vector $(\bb X,Y)$. The main objective is usually to predict the conditional mean $\mathbb E(Y\mid \bb X = \bb x)$, which corresponds to minimizing the squared error prediction loss. While the mean summarizes the behavior of $Y$ in the center of its distribution, applications in the field of risk assessment require knowledge of the distributional tail. For a probability level $\tau \in (0,1)$, an important quantity is the conditional quantile
\begin{align}\label{eq:cond_quant}
  Q_{\bb x}(\tau) = F_Y^{-1}(\tau \mid \bb X=\bb x),
\end{align}
where $F_Y^{-1}(\cdot\mid \bb X=\bb x)$ is the generalized inverse of the conditional distribution function of $Y$ given $\bb X=\bb x$.
There has been extensive research in statistics and machine learning to adapt mean prediction methods  to other loss functions than squared error. For instance, quantile regression
relies on minimizing the conditional quantile loss, which is based on the quantile check function \citep{Koenker1978}. This has been extended to more flexible regression functions such as the quantile regression forest \citep{Meinshausen2006} and the gradient forest \citep{Athey2019}, which both build on the original random forest \citep{Breiman2001}. Another popular tree-based method in machine learning is gradient boosting by \cite{Friedman2001}. This versatile method aims at optimizing an objective function with a recursive procedure akin to gradient descent.

Let $n$ denote the sample size and $\tau=\tau_n$ the quantile level. The existing quantile regression methodology works well in the case of a fixed quantile level, or in the case of a quantile that is only moderately high, that is, $\tau_n \to 1$ and $n(1-\tau_n)\to \infty$ as $n\to \infty$, meaning that there are sufficient observations above the $\tau_n$ level. For more extreme quantiles with  $n(1-\tau_n) \to c \in [0,\infty)$, the quantile loss function is no longer useful because observations become scarce at that level and extrapolation beyond the range of observed values is needed.
Extreme value theory provides the statistical tools for a sensible extrapolation into the
tail of the variable of interest $Y$. For a large threshold $u$ close to
the upper endpoint of the distribution of $Y$, the distribution of the threshold exceedance $Y - u \mid Y > u$
can be approximated by the generalized Pareto distribution (GPD)
\begin{equation}\label{eq:gpd}
  H_{\gamma,\sigma}(y)= 1-\left(1+\gamma y/\sigma \right)_+^{-1/\gamma},\quad y\geq 0,
\end{equation}
where for $a\in \mathbb R$, $a_+ = \max(0,a)$, and $\gamma\in\mathbb{R}$ and $\sigma>0$ are the shape and scale parameters, respectively.

\tcb{There are three main streams in the literature focusing on the estimation of covariate dependent extreme quantiles. First, a parametric form (e.g. linear) can be assumed for the conditional quantile function  \eqref{eq:cond_quant} and  estimators for extreme quantiles can be derived and  studied as in \cite{C05}. The second stream uses GPD modeling of exceedances above a high threshold and assumes that the parameters $\sigma(\bb x)$ and $\gamma(\bb x)$  depend on the covariates via  parametric or semi-parametric models \citep{DS90, WangTsai2009} or generalized additive models \citep{CDD05, you2019}. The third stream is a fully non-parametric local approach where local smoothing estimation techniques for the conditional quantile at moderately high levels are applied and then extrapolated to the extreme level. For example, \cite{Daouia2013} and \cite{Velthoenetal2019} apply kernel smooting estimation for the conditional tail distribution and the conditional quantile, respectively, and \citep{GardesStupfler2019} considers a covariate dependent adaption of Weissman's estimation for heavy-tailed data.}
While linear or additive models are restricted in their modeling flexibility, local smoothing methods are known to be sensitive to the curse of dimensionality and work well only for a low-dimensional predictor space.  To bypass these issues for modern applications with complex data, tree-based methods are attractive due to  their modelling flexibility and robustness in higher dimensions. A first contribution to the use of tree-based models in extreme value theory  is the generalized Pareto regression tree  \cite{FLT19}, but a single tree is used resulting with a model with limited flexibility and predictive performance.

Our goal is to estimate the extreme conditional quantile $Q_{\bb x}(\tau)$ in  \eqref{eq:cond_quant}, where the dimension of covariates  $d$ is large and the response surface allows for complex non-linear effects. To this end, we build a bridge between the predictive power of tree-based ensemble methods from machine learning and the theory of extrapolation from extreme value theory. Following the second stream of research mentioned above, we model the tail of the conditional  distribution of $Y$ given $\bb X=\bb x$ using  a GPD distribution in \eqref{eq:gpd} with covariate dependent parameters $\gamma(\bb x)$ and $\sigma(\bb x)$. We propose \texttt{gbex}, a gradient boosting algorithm to optimize the deviance (negative log-likelihood) of the GPD model, to estimate $\gamma(\bb x)$ and $\sigma(\bb x)$. 
In each boosting iteration, these parameters are updated based on an approximation of the deviance gradient by regression trees. The resulting model includes many trees and is flexible enough to account for a complex non-linear response surface. The boosting algorithm has several tuning parameters, the most important ones being the number of  trees and the tree depth.  We show how they can be chosen effectively using cross-validation. 

In two numerical experiments we illustrate that, for the task of extremal quantile estimation, our methodology outperforms  quantile regression approaches that do not use tail extrapolation \citep{Meinshausen2006, Athey2019} and methods from extreme value theory that assume simple forms  for $\gamma(\bb x)$ and $\sigma(\bb x)$ such as generalized additive models \citep{you2019}. As a result, to the best of our knowledge, our gradient boosting is the first method that reliably estimates extreme quantiles in the case of complex predictor spaces and in the presence of possibly high-dimensional noise variables.

We apply the developed method to forecast the extreme quantiles of daily precipitation in the Netherlands using the output of numerical weather prediction models as covariates. Our diagnostic tools, namely variable importance score and partial dependence plots, are able to identify changes in the tail heaviness of precipitation as seasonality patterns in the shape parameter estimates $\gamma(\bb x)$. We further investigate  the contribution of weather prediction model outputs of neighbouring stations to forecasting the extreme precipitation of a specific location.

\tcb{Our main contribution is methodological and demonstrates that the tree-based modeling of extremes initiated in \cite{FLT19} with a single tree can be extended to a powerful ensemble method thanks to boosting. Algorithm~\ref{algo:gbex} is an adaptation of Friedman's gradient boosting \cite{Friedman2001,Friedman2002} to the GPD model, with an extra clipping gradient step introduced for numerical stability. The  methodology and resulting algorithm are explained in detail for the sake of completeness and pedagogy. Beyond this, the overall procedure in Algorithm~\ref{algo:gbex2} combines extreme value theory and machine learning in two ways: we propose an adaptive covariate dependent threshold to define the exceedances that are the input of Algorithm~\ref{algo:gbex} and we introduce the extreme conditional quantile estimator. A general issue with gradient boosting is selection of hyperparameter such as the number of trees and the parameters governing the tree structure. We propose adaptive hyperparameters selection with deviance-based cross-validation. This is not straightforward since the quantity of interest, namely, the extreme conditional quantile, has no clear relationship with deviance. Our choice is due to the fact that the more natural pinball loss used in quantile regression degenerates in the extreme regime $\tau_n\to 0$. Our simulation studies reveals that deviance-based cross-validation  performs well also for extreme quantile estimation (see Figure~\ref{fig: B_depth_CV}).
The asymptotic analysis of our \texttt{gbex} algorithm is challenging and beyond the scope of this paper, mainly due to two issues: the GPD deviance is non convex while all of the existing  theory on gradient boosting considers convex loss functions;  model misspecification has to be taken into account since the GPD model is only an approximation for the threshold exceedances.}
 
\tcb{There has been active research on machine learning methods for extremes in parallel to this paper. Extremal random forests \cite{gne2022} are another proposal for tree-based GPD modelling where the localizing weights of a generalized random forest \cite{Athey2019} are used. Extreme quantile regression via neural networks is considered in \cite{pas2022,ric2022}. \cite{gne2022} and \cite{pas2022} provide comparative simulation studies of the different approaches. As pointed out by a referee, another line of research for extremes in complex high-dimensional models consists in dimension reduction techniques as in the single index model for  extreme quantile estimation \citep{Gardes2019}.}

 The paper is organized as follows. Section~\ref{sec:methodology} introduces our methodology and algorithms for extreme quantile regression based on GPD modeling with gradient boosting.  Practical questions such as parameter tuning and model interpretation are discussed in Section~\ref{sec:tuning-parameters}, while Section~\ref{sec:num-exp} is devoted to assessing the performance of our method in two simulation studies. The application to statistical post-processing of weather forecasts with precipitation data in the Netherlands is given in Section~\ref{sec:application}. We conclude the paper with a summary and a discussion of future research directions. 

The gradient boosting method is implemented in an R package and can be downloaded from GitHub at \url{https://github.com/JVelthoen/gbex/}

\section{Extreme quantile regression with gradient boosting}\label{sec:methodology}
\subsection{Background on extreme quantile estimation} \label{sec:model}
Extreme value theory provides the asymptotic results for extrapolating beyond the range of the data and statistical methodology has been developed to accurately estimate extreme quantiles.
\tcb{In the simplest case with no covariate, a sample of $n$ independent copies $Y_1,\dots, Y_n$ of the response $Y$ is observed and the goal is to estimate a quantile $Q(\tau_n)$ of $Y$ at an extreme probability level $\tau_n \in (0,1)$. Here, extreme means that $\tau_n \to 1$ and $n(1-\tau_n) \to c \geq 0$ as $n \to \infty$, that is, the expected number of observations that exceed $Q(\tau_n)$ does not go to infinity as $n\to \infty$.
Empirical estimation then becomes strongly biased and extrapolation beyond observations is needed.
The usual strategy is to use the empirical quantile $Y_{k-n:n}$ as a threshold and to consider exceedances above this threshold. Asymptotic theory assumes that $k=k(n)$, the number of observations above threshold satisfies $k\to\infty$ and $k/n\to 0$. Stated differently, $Y_{k-n:n}$ is the empirical quantile at level $\tau_{0,n}=1-k/n$. The level $\tau_{0,n}$ is said intermediate as it satisfies $\tau_{0,n} \to 1$ and $n(1-\tau_{0,n})\to \infty$. These distinctions are particularly important for the asymptotic theory of estimators in extreme value theory \cite{deh2006a}.}

One of the main results for extrapolation in the univariate case is the Pickands--de Haan--Balkema theorem \citep{bal1974, pic1975}, which states that under mild regularity conditions on the tail of the distribution of $Y$, the rescaled distribution of exceedances over a high threshold converges to the generalized Pareto distribution. More precisely, if $y^*$ denotes the upper endpoint of the distribution of $Y$ then there exist a normalizing function $\sigma(u) > 0$ such that 
\begin{equation}\label{eq:rescaled-exceedances}
\lim_{u\uparrow y^*}\mathbb{P}\left(\frac{Y-u}{\sigma(u)}>y \, \mid \,Y>u \right) =1-H_{\gamma, 1}(y), \quad y \geq 0,
\end{equation}
where $H$ is  defined in~\eqref{eq:gpd}, with the convention $H_{0, \sigma}(y)=1-\exp(-y/\sigma)$, $y\geq 0$.
The shape parameter $\gamma \in \mathbb R$ indicates the heaviness of the upper tail of $Y$, where
$\gamma < 0$, $\gamma = 0$ and $\gamma >0$ correspond to distributions respectively with short tails (e.g., uniform), light tails (e.g., Gaussian, exponential) and  power tails (e.g., Student's $t$). 

Moreover, the GPD is the only non-degenerate distribution that can arise as the limit of threshold exceedances as in \eqref{eq:rescaled-exceedances}, and therefore it is an asymptotically motivated model for tail extrapolation and high quantile estimation. By the limit relation in \eqref{eq:rescaled-exceedances}, for a large threshold $u$, the conditional distribution of $Y - u$ given $Y> u$ can be approximated by  $H_{\gamma,\sigma}$ with $\sigma = \sigma(u)$. The threshold $u$ can be chosen as the quantile $Q(\tau_0)$ of $Y$ for some intermediate probability level $\tau_0 \in (0,1)$.
Inverting the distribution function in \eqref{eq:gpd} provides an approximation of the quantile for probability level $\tau>\tau_0$ by
\begin{equation}\label{eq:extreme-quantile}
	Q(\tau) \approx Q(\tau_0) + \sigma \frac{\left( \frac{1-\tau}{1-\tau_0} \right)^{-\gamma}-1}{\gamma}.
\end{equation}

\subsection{Setup for extreme quantile regression}
We consider here the setting where the response $Y_i\in \RR$ depends on covariates $\bb X_i\in \RR^d$ and our goal is to develop an estimator for the conditional quantile $Q_{\bb x}(\tau)$ defined by \eqref{eq:cond_quant} at an extreme quantile level $\tau=\tau_n$. \tcb{For this purpose, exceedances above an intermediate quantile $\tau_0=\tau_{0,n}$ will be considered; see the beginning of Section~\ref{sec:model} for a discussion on extreme and intermediate quantiles}. Recall that $(\bb X_1,Y_1),\dots ,(\bb X_n,Y_n)$ denote  independent copies of the random vector $(\bb X,Y)$ with $\bb X \in \RR^d$ and $Y \in \RR$.

In this setup, the intermediate threshold $Q(\tau_0)$, shape parameter $\gamma$ and scale parameter $\sigma$ in~\eqref{eq:extreme-quantile} may depend on covariates. \tcb{We therefore assume that the GPD approximation in~\eqref{eq:rescaled-exceedances} holds pointwise for any $\bb x \in \mathbb R^d$ with $u(\bb x) = Q_{\bb x}(\tau_0)$, $\gamma(\bb x)$ and $\sigma(\bb x)$, where for the scale we omit the dependence on the intermediate level $u(\bb x)$ for simplicity. The approximation for the extreme conditional quantile becomes}
\begin{equation} \label{eq:conditional-extreme-quantile}
Q_{\bb x}(\tau)\approx Q_{\bb x}(\tau_0)+\sigma(\bb x) \frac{\left( \frac{1-\tau}{1-\tau_0}\right)^{- \gamma(\bb x)}-1}{ \gamma(\bb x)},\quad \tau>\tau_0.
\end{equation}
The triple $(Q_{\bb x}(\tau_0),\sigma(\bb x),\gamma(\bb x))$ provides a model for the tail (that is  above the probability level $\tau_0$) of the conditional law of $Y$ given $\bb X=\bb x$. An estimator of conditional extreme quantiles $\hat{Q}_{\bb x}(\tau)$ is obtained by plugging in estimators $(\hat{Q}_{\bb x}(\tau_0),\hat{\sigma}(\bb x),\hat{\gamma}(\bb x))$ in Equation \eqref{eq:conditional-extreme-quantile}.

In the following we propose estimators for these three quantities.  Our main contribution is a gradient boosting procedure for estimation of the GPD parameters $(\sigma(\bb x),\gamma(\bb x))$ that allows flexible regression functions with possibly many covariates. For estimation of the intermediate quantile $Q_{\bb x}(\tau_0)$, any method for (non-extreme) quantile regression can be used and we outline in Section~\ref{sec:threshold} how the existing method of quantile random forests can be applied.

\subsection{GPD modeling with gradient boosting} \label{sec:grad-boosting}
\tcb{In this section we propose the \texttt{gbex} algorithm to estimate the GPD parameters $(\sigma(\bb x),\gamma(\bb x))$ using gradient boosting to build an ensemble of tree predictors. The  algorithm is the standard Friedman's boosting algorithm \cite{Friedman2001,Friedman2002} applied with objective function given by the GPD negative log-likelihood (also called deviance in the machine learning literature). Since the GPD has two parameters, two sequences of trees are needed; this is similar to the strategy for multiclass classification where several sequences of trees are trained to learn the different class probabilities.}

Based on Equation~\eqref{eq:rescaled-exceedances}, the Peaks-over-Threshold approach assumes that, given $\bb X=\bb x$, the excess of $Y$ above the threshold $Q_{\bb x}(\tau_0)$ follows approximately a GPD.  \tcb{In order to compute the sample of exceedances, we rely on a (non-extreme) quantile regression method providing an estimation of the intermediate quantile function $\bb x \mapsto \hat Q_{\bb x}(\tau_0)$. Applying this estimated function at the predictor values $\bb X_1,\dots ,\bb X_n$, we obtain the exceedances above the intermediate threshold defined as}
  \begin{align}\label{def_exc}
    Z_i = \left(Y_i - \hat{Q}_{\bb X_i}(\tau_0)\right)_+, \quad i=1,\dots, n.
    \end{align}
  Note that $Z_i=0$ whenever the value $Y_i$ is below  threshold. We assume that the  intermediate threshold is high enough so that the exceedances  can be modeled by the generalized Pareto distribution and the approximation of conditional quantiles \eqref{eq:conditional-extreme-quantile} is good. Our aim is to learn the conditional parameter $\theta(\bb x)=(\sigma(\bb x),\gamma(\bb x))$ based on the sample of exceedances above the threshold. We apply tree-base gradient boosting  \cite{Friedman2001, Friedman2002} and use the GPD deviance (negative log-likelihood) as the objective function to minimize.

In absence of covariates, a standard way of estimating the GPD parameters $\theta = (\sigma, \gamma)$  is the maximum likelihood method \citep{Smith87}, which provides asymptotically normal estimators in the unconditional case with $\gamma>-1/2$. The negative log-likelihood, or deviance, for an exceedance $Z_i$ from a GPD distribution with parameters $\theta(\bb X_i) = (\sigma(\bb X_i),\gamma(\bb X_i))$ is given by

\begin{equation}\label{eq:likelihood}
\ell_{Z_i}(\theta(\bb X_i) )= \left[(1+1/\gamma(\bb X_i) )\log\left(1+\gamma(\bb X_i)  \frac{Z_i}{\sigma(\bb X_i) }\right) + \log \sigma(\bb X_i) \right]\mathds{1}_{Z_i>0}.
\end{equation}

\tcb{The gradient boosting algorithm starts with an initial estimate, which is given by the unconditional maximum likelihood estimator, that is
\begin{equation}\label{eq:theta-init}
	\theta_0(\bb x)\equiv \theta_0 = \argmin_{\theta} \sum_{i=1}^n\ell_{Z_i}(\theta).
\end{equation}
This initially constant model is then gradually improved in an additive way.
The big picture is the following. Starting from the initial constant model $\theta_0=(\sigma_0,\gamma_0)$, we sequentially construct a sequence of $B$ pairs of trees.
At step $b$, the goal is to improve the current model  $\theta_{b-1}(\bb x)=(\sigma_{b-1}(\bb x),\gamma_{b-1}(\bb x))$ by  adding a pair of gradient trees $(T_b^\sigma(\bb x), T_b^\gamma(\bb x))$. For $b=1,\ldots,B$:
\begin{itemize}
\item[i)] a subsample is randomly drawn from the set of exceedances $(\bb X_i,Z_i)_{1\leq i\leq n}$;
\item[ii)] on this subsample, the residuals (deviance derivatives) with respect to $\sigma$ and $\gamma$ are computed;
\item[iii)] two regression trees are fitted on these residuals; they provide two partitions of the feature space into different leaves; on the different leaves, the tree values are modified by line search approximation so as to minimize the deviance;
\item[iv)] the model is updated by adding a shrunken version of these trees.
\end{itemize}
}
\tcb{We now provide mathematical details for each of the different steps. We assume that the reader is familiar with the standard regression tree based on square loss minimization (CART algorithm as in \cite{BFSO84} or \cite{ESL}). For $b=1,\ldots,B$:
\begin{itemize}
\item[i)] A random subset  $S_b \subset \{1, \ldots , n\}$ of size $[sn]$ is randomly drawn, where the parameter $s\in (0,1]$ is called the \textit{subsampling fraction}.
\item[ii)] The model residuals are computed on the subsample $S_b$ by
\begin{equation*} 
	r_{b,i}^{\sigma} =\frac{\partial \ell_{Z_i}}{\partial \sigma}(\theta_{b-1}(\bb X_i))\quad \mbox{and}\quad 	r_{b,i}^{\gamma} =\frac{\partial \ell_{Z_i}}{\partial \gamma}(\theta_{b-1}(\bb X_i)), \quad i \in S_b.
\end{equation*}
The deviance derivatives are provided in Appendix~\ref{app:derivatives}.
\item[iii)] A pair of regression trees $(T_b^\sigma(\bb x), T_b^\gamma(\bb x))$ are fitted, respectively on $(\bb X_i, r_{b,i}^\sigma)_{i\in S_b}$ and $(\bb X_i, r_{b,i}^\gamma)_{i\in S_b}$. The tree construction uses the standard CART algorithm \cite{BFSO84} based on recursive binary splitting and provides a partition of the feature space into several rectangles called leaves. Several parameters are involved in the stopping rule: the maximal depths $D^\sigma,D^\gamma$ (i.e., the maximum number of splits between the root and a leaf in the tree) and the minimal leaf sizes $L_{\min}^\sigma,L_{\min}^\gamma$ (minimum number of observations in each leaf).  The leaves of $T_b^\sigma$ (resp. $T_b^\gamma$) are denoted by $(L^{\sigma}_{b,j})_{1\leq j\leq J_b^\sigma}$ (resp. $(L^{\sigma}_{b,j})_{1\leq j\leq J_b^\sigma}$).
Following \cite{Friedman2001}, the regression trees are then modified: the partitions are kept unchanged but the tree values are chosen so as to minimize the deviance. This is done by line search, that is, the updated value $\xi_{b,j}^\sigma$ in leaf  $L^{\sigma}_{b,j}$ is obtained by solving the minimization problem
\begin{equation} \label{eq:linesearch}
	\xi_{b,j}^\sigma=\argmin_{\xi} \sum_{\bb X_i\in L_{b,j}^\sigma}\ell_{Z_i}(\theta_{b-1}(\bb X_i)+\xi e_\sigma),\quad j=1,\ldots,J_b^\sigma,
\end{equation}
where $e_\sigma=(1,0)$ gives the direction of the line search corresponding to $\sigma$. For the parameter $\gamma$, the line search is performed  in direction $e_\gamma=(0,1)$, yielding the value $\xi_{b,j}^\gamma$  in the leaf $L^{\gamma}_{b,j}$  (same equation with $\sigma$ replaced by $\gamma$ everywhere). In practice the line search~\eqref{eq:linesearch} can be computationally expensive and an approximation is used instead, i.e., a Newton--Raphson step resulting in
\begin{equation*} 
	\tilde \xi_{b,j}^\sigma =-\dfrac{\sum_{\bb X_i\in L_{b,j}^\sigma} \dfrac{\partial \ell_{Z_i}}{\partial\sigma}(\theta_{b-1}(\bb X_i))}{\sum_{\bb X_i\in L_{b,j}^\sigma} \dfrac{\partial^2 \ell_{Z_i}}{\partial \sigma^2}(\theta_{b-1}(\bb X_i))}.
\end{equation*}
For the parameter $\gamma$,  the line search approximation in leaf $L_{b,j}^\gamma$ yields the value $\tilde \xi_{b,j}^\gamma$ (same equation with $\sigma$ replaced by $\gamma$ everywhere). The gradient trees are  given by
\begin{equation}\label{eq:tree-0}
T_b^\sigma(\bb x)=\sum_{j=1}^{J_b^\sigma} \tilde{\xi}_{b,j}^\sigma  \mathds{1}_{\{\bb x\in L_{b,j}^\sigma\}} \quad\mbox{and}\quad 
T_b^\gamma(\bb x)=\sum_{j=1}^{J_b^\gamma} \tilde{\xi}_{b,j}^\gamma  \mathds{1}_{\{\bb x\in L_{b,j}^\gamma\}} .
\end{equation}
\item[iv)] The model $\theta_{b-1}(\bb x)=(\sigma_{b-1}(\bb x),\gamma_{b-1}(\bb x))$ is finally updated by
\begin{align}
\theta_b(\bb x)&=(\sigma_{b}(\bb x),\gamma_{b}(\bb x))\nonumber\\
&=(\sigma_{b-1}(\bb x)+\lambda^\sigma T_b^\sigma(\bb x), \gamma_{b-1}(\bb x)+\lambda^\gamma T_b^\gamma(\bb x)),\label{eq:update}
\end{align}
where the shrinkage parameters $\lambda^{\sigma},\lambda^\gamma\in (0,1)$ are called  learning rates. They are used to slow down the dynamics since only a shrunken version of the trees is added to the current model. 
\end{itemize}
}

\tcb{The final output for the estimated parameters is the  gradient boosting model 
\begin{equation}\label{eq:estimator}
	\hat\sigma(\bb x)=\sigma_0+\lambda^\sigma\sum_{b=1}^B T_b^\sigma(\bb x), \quad 
	\hat\gamma(\bb x)=\gamma_0+\lambda^\gamma\sum_{b=1}^B T_b^\gamma(\bb x).
\end{equation}
The algorithm as described above is an adaptation of Friedman's gradient boosting algorithm \cite{Friedman2001, Friedman2002} to the conditional GPD model for exceedances above a threshold. In a first implementation of this algorithm, we could observe a numerical instability due to the fact that the GPD negative log-likelihood is not Lipschitz and that its gradient may explode. For this reason, we introduce \textit{gradient clipping}, a standard trick in machine learning to avoid gradient explosion \citep{pmlr-v130, pmlr-v139}. This means that we bound the absolute value of the Newton--Raphson step by~$1$ in order to mitigate the strong influence of extreme observations, leading to 
\begin{equation}\label{eq:tree}
T_b^\sigma(\bb x)=\sum_{j=1}^{J_b^\sigma}  \mbox{sign}(\tilde{\xi}_{b,j}^\sigma) \min(\vert\tilde \xi_{b,j}^\sigma\vert,1)  \mathds{1}_{\{\bb x\in L_{b,j}^\sigma\}} 
\end{equation}
and similarly for $T_b^\gamma(\bb x)$. We observe in practice that gradient clipping results in a more stable algorithm with better performance.
}

Algorithm~\ref{algo:gbex} summarizes the procedure for GPD modeling of exceedances. In practice, the number of iterations $B$ is an important parameter and its choice corresponds to a trade-off between  bias and variance. The procedure is  prone to overfitting as $B \to \infty$ and cross-validation is used to prevent this by early stopping; see Section~\ref{sec:tuning} where we discuss the interpretation of the different tuning parameters and their selection in practice. 

\begin{algorithm}[H]
\caption{gbex boosting algorithm for GPD modeling}
\label{algo:gbex}
{\small
\begin{flushleft}
\textbf{Input:} 
\begin{itemize}
\item $\theta_0$: the initial values of the parameters with default value as in \eqref{eq:theta-init};
\item $(\bb X_i,Z_i)_{1\leq i\leq n}$: data sample of exceedances above threshold; 
\item $B$: number of gradient trees;
\item $D^\sigma$, $D^\gamma$: maximum tree depth for the gradient trees; 
\item $\lambda^\sigma,\lambda^\gamma$: learning rates for the update of the GPD parameters $\sigma$ and $\gamma$ respectively; 
\item $s$: subsampling fraction; 
\item $L_{\min}^{\sigma}$, $L_{\min}^\gamma$: minimum leaf size of the  nodes in the trees.
\end{itemize}
\textbf{Algorithm:}
For $b=1,\ldots,B$:
\begin{enumerate}
\item Draw a random subsample $S_b\subset\{1,\ldots,n\}$ of size $[s n]$.
\item Compute the deviance derivatives on the subsample $S_b$:
\[
r_{b,i}^\sigma=	\frac{\partial \ell_{Z_i}}{\partial \sigma}(\theta_{b-1}(\bb X_i))\quad \mbox{and}\quad
r_{b,i}^\gamma=	\frac{\partial \ell_{Z_i}}{\partial \gamma}(\theta_{b-1}(\bb X_i)),
\quad i\in S_b.
\]
\item Fit regression trees $T^\sigma_b$, $T^\gamma_b$ that predict the gradients $r_{b,i}^\sigma$ and $r_{b,i}^\gamma$ as functions of the covariates $\bb X_i$ on the sample $i\in S_b$; the trees are built with maximal depth $(D^\sigma,D^\gamma)$ and  minimal leaf size $(L_{\min}^\sigma, L_{\min}^\gamma )$; for the tree values, use the line search approximation with gradient clipping \eqref{eq:tree}.
\item Update the GPD parameters $\theta_{b}(\bb x)= (\hat \sigma_{b}(\bb x),\hat \gamma_{b}(\bb x))$ with learning rates $(\lambda^\sigma,\lambda^\gamma)$, i.e., 
\[
\hat \sigma_{b}(\bb x)=\hat \sigma_{b-1}(\bb x)+\lambda^\sigma T_b^\sigma(\bb x)
\quad\mbox{and}\quad 
\hat \gamma_{b}(\bb x)=\hat \gamma_{b-1}(\bb x)+\lambda^\gamma T_b^\gamma(\bb x).
\]
\end{enumerate}
\textbf{Output:} Conditional GPD parameters $(\hat\sigma(\bb x),\hat\gamma(\bb x))=(\hat\sigma_B(\bb x),\hat\gamma_B(\bb x))$.
\end{flushleft}
}
\end{algorithm}

\subsection{Extreme quantile regression}\label{sec:threshold}

The input of Algorithm~\ref{algo:gbex} is the sample of exceedances $Z_i$ defined by~\eqref{def_exc}. The conditional intermediate quantile $\hat{Q}_{\bb X_i}(\tau_0)$ used in this definition generally also depends on the covariate vector $\bb X_i$ and needs to be modeled first. For this task, any method for (non-extreme) quantile regression can be used, but we note that the quality of the approximation~\eqref{eq:conditional-extreme-quantile} of the extreme quantile will also depend on the accuracy of the intermediate quantile estimate. Together with the gradient boosting procedure for the GPD parameters in Section~\ref{sec:grad-boosting}, we obtain an algorithm for extreme quantile prediction.
We refer to this algorithm as the \texttt{gbex} method. It combines the flexibility of gradient boosting with the extrapolation technique from extreme value theory. 

\begin{algorithm}%[H]
  \caption{gbex algorithm for extreme quantile prediction}
  \label{algo:gbex2}
{\small
\begin{flushleft}
\textbf{Input:} 
\begin{itemize}
\item $(\bb X_i,Y_i)_{1\leq i\leq n}$: data sample; 
\item $\tau_0$:  probability level for the threshold;
\item $\tau$: probability level for the prediction such that $\tau>\tau_0$;
\item parameters of the gbex boosting algorithm for GPD modeling of exceedances (Algorithm~\ref{algo:gbex}).
\end{itemize}
\textbf{Algorithm:}
\begin{enumerate}
\item Fit a quantile regression to the sample $(\bb X_i,Y_i)_{1\leq i\leq n}$ that provides estimates $\hat Q_{\bb x}(\tau_0)$ of the conditional quantiles of order $\tau_0$.
\item Compute the exceedances $Z_i=(Y_i-\hat Q_{\bb X_i}(\tau_0))_+$, $1\leq i\leq n$.
\item Let $I=\{i:Z_i>0\}$ be the index set of positive exceedances and run Algorithm~\ref{algo:gbex} on the data set $(\bb X_i,Z_i)_{i\in I}$ to estimate the GPD parameters $(\hat \sigma(\bb x), \hat\gamma(\bb x))$.
\end{enumerate}
\textbf{Output:} Estimation of the extreme conditional quantile
\[
\hat Q_{\bb x}(\tau)= \hat Q_{\bb x}(\tau_0)+\hat\sigma(\bb x) \frac{\left( \frac{1-\tau}{1-\tau_0}\right)^{- \hat\gamma(\bb x)}-1}{ \hat\gamma(\bb x)}.
\]
\end{flushleft}
}
\end{algorithm}

While in principle any quantile regression method can be used for estimation of the conditional intermediate quantiles $\hat Q_{\bb X_i}(\tau_0)$, we propose to use a quantile random forest. The reason for this is three-fold: first it requires no parametric assumptions on the quantile functions; secondly it exhibits good performance for high dimensional predictor spaces; finally it requires minimal tuning for good results.
Quantile regression forests were first proposed by \cite{Meinshausen2006} using the weights from a standard random forest \citep{Breiman2001}. The drawback of this method is that the criterion used in recursive binary splitting to build the trees of the random forest is not tailored to quantile regression.  \cite{Wager2018} therefore define a generalized random forest with splitting rule designed for that specific task, where the splitting criterion is related to the quantile loss function. 
In our case, we require the estimator of $Q_{\bb x}(\tau_0)$ at the sample points $\bb x \in\{\bb X_1,\ldots,\bb X_n\}$ and we recommend the use of out-of-bag estimation
$ \hat Q_{\bb X_i}(\tau_0) =  \hat{Q}_{\bb X_i}^{oob}(\tau_0)$.
This means that only the trees for which the $i$th observation is out-of-bag are kept for the quantile estimation  at $\bb x=\bb X_i$, that is, trees based on sub-samples not containing the $i$th observation. This is necessary to avoid giving too much weight to the $i$th observation when predicting  at $\bb x=\bb X_i$.

\section{Parameter tuning and interpretation}\label{sec:tuning-parameters}

\subsection{Parameter tuning}\label{sec:tuning}

Our gradient boosting procedure for GPD modelling includes several  parameters that need to be tuned properly for good results. We discuss in this section the interpretation of the different parameters and how to choose them. We introduce  data driven choices based on cross validation  for the most sensitive parameters and suggest sensible default values for the remaining parameters. \tcb{This concerns the tuning parameters of Algorithm~\ref{algo:gbex} that takes the sample of exceedances as input and we therefore consider cross-validation within the sample of exceedances.}

\subsubsection{Tree number $B$} The number of trees is the most important regularization parameter. The boosting procedure starts from a constant model, that is usually an underfit, and adds recursively trees that adapt the model to the data, leading eventually to an overfit. 

We recommend repeated $K$-fold cross-validation based on the deviance for a data driven choice of $B$. Given a maximal tree number $B_{max}$ and a division of the data set into $K$ folds $\mathcal{D}_1, \dots, \mathcal{D}_K$, we repeatedly run the algorithm with $B_{max}$ iterations on the data with one fold left-out and then compute the deviance on the left-out fold as a function of  $B$. Adding up the deviances for the different folds, we obtain the cross-validation deviance. More formally, we define
\begin{equation} \label{eq: CV_D}
		\mathrm{DEV}_{CV}(B) = \sum_{k=1}^K \sum_{i \in \mathcal{D}_k} \ell_{Z_i}(\hat{\theta}_B^{-\mathcal{D}_k}(\bb X_i)),  \quad B=0,\ldots,B_{max},
              \end{equation}
where $\hat \theta_B^{-\mathcal{D}_k}$ denotes the model with $B$ trees trained on the data sample with the $k$th fold $\mathcal{D}_k$ held out. Due to large values of the deviance on extreme observations, the cross-validation deviance is prone to fluctuations with respect to the partition into folds and we therefore recommend repeated cross-validation. A typical choice is $K=5$ or $10$ with $5$ repetitions.  The choice of $B$ is then the minimizer of the cross-validation deviance.

\subsubsection{Tree depth $(D^\sigma,D^\gamma)$}\label{sec:tuning_depth}
The gradient boosting algorithm outputs a sum of tree functions. The complexity of the model is therefore determined by the depth parameters $D^{\sigma}$ and $D^{\gamma}$, also called interaction depths. A zero depth tree corresponds to a constant tree with no split, so that $D^\sigma=0$ or $D^\gamma=0$ yield models with constant scale or shape parameters, respectively. Since the extreme value index $\gamma$ is notoriously difficult to estimate, it is common in extreme value theory to assume a constant value $\gamma(\bb x)\equiv \gamma$ so that the case $D^\gamma=0$ is particularly important. A tree with depth~$1$, also called a stump, makes only one single split on a single variable. As a result, $D^\sigma=1$ (resp.~$D^\gamma=1$) corresponds to an additive model in the predictors for  $\sigma(\bb x)$ (resp.~$\gamma(\bb x)$).  Trees with larger depth allow to introduce interaction effects between the predictors of order equal to the depth parameter. \tcb{We refer to  \cite[Section 10.11]{ESL} for a more detailed discussion on tree depth and interaction order in gradient boosting.}

\begin{figure}[hbt!]
    \centering
    \begin{minipage}{.5\textwidth}
        \centering
        \includegraphics[width=0.9\linewidth]{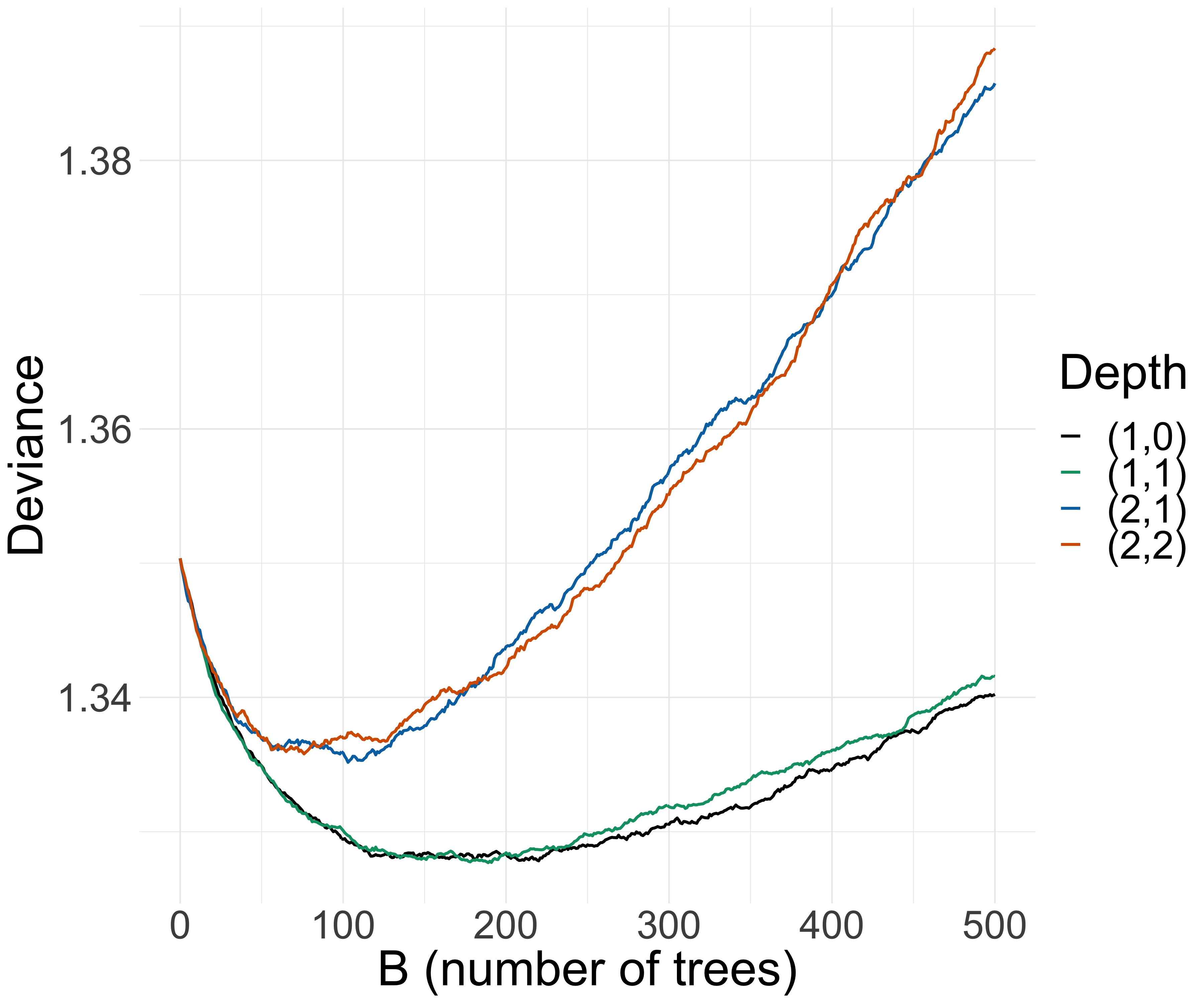}
      \end{minipage}%
    \begin{minipage}{0.5\textwidth}
        \centering
        \includegraphics[width=0.9\linewidth]{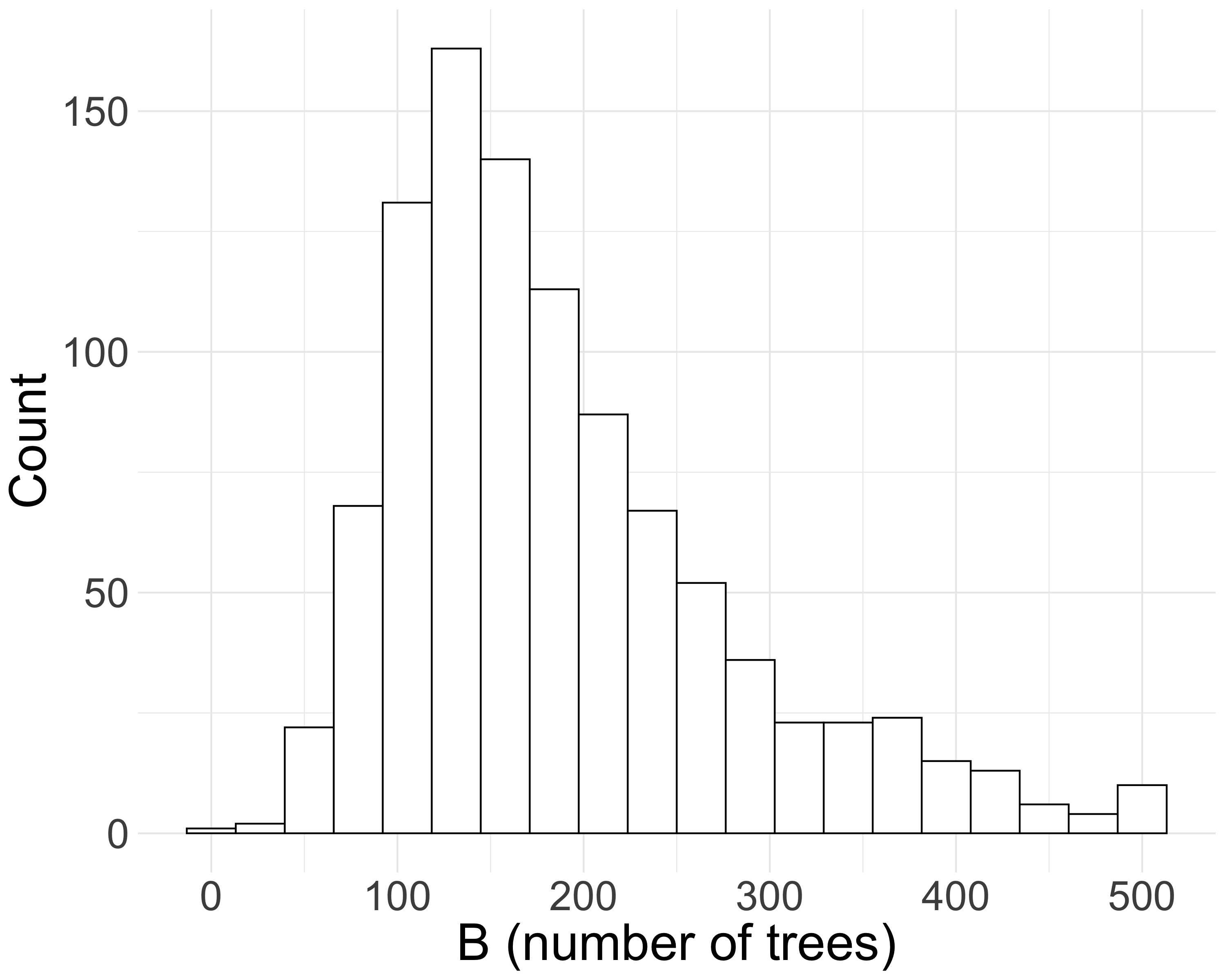}
    \end{minipage}
    \caption{Left panel: cross-validation deviance given by \eqref{eq: CV_D} against $B$ for one random sample and depth $(D^\sigma,D^\gamma)=(1,0)$, $(1,1)$, $(2,1)$ and $(2,2)$. Right panel: selected values of $B$ for $1000$ samples when $(D^\sigma,D^\gamma)=(1,0)$ is fixed. The design of the simulation study is  Model 1 described in Section~\ref{sec:num-exp}.}
    \label{fig:CV}
\end{figure}

In practice,  the depth parameter is quite hard to tune and we recommend to consider depth no larger than $3$, also because interactions of higher order are difficult to interpret. Based on our experience, sensible default values are $D^\sigma=2$ and  $D^\gamma=1$. But more interestingly, cross-validation can be used to select the depth parameters. The left panel of Figure~\ref{fig:CV} shows a typical cross-validation diagnostic  in the context of the simulation study detailed in Section~\ref{sec:num-exp}. Here $B_{max}=500$ and depths parameter $(D^\sigma,D^\gamma)=(1,0)$, $(1,1)$, $(2,1)$ and $(2,2)$ are considered. The plot shows that sensible choices are $B\approx 200$ and $(D^\sigma,D^\gamma)=(1,0)$ or $(1,1)$ (more details given in Section~\ref{sec:num-exp}). The histogram in the right panel shows that, depending on the randomly simulated sample, $B$  typically lies in the range $[100,250]$, where the deviance is relatively flat ($(D^\sigma,D^\gamma)=(1,0)$ is fixed here).

\subsubsection{Learning rates $(\lambda^\sigma,\lambda^\gamma)$} As usual in gradient boosting, there is a balance between the learning rate and the number of trees. As noted in  \cite{Ridgeway07}, multiplying the learning rate by $0.1$ roughly requires $10$ times more trees for a similar result. It is common to fix the learning rate to a small value, typically $0.01$ or $0.001$, and to consider the tree number as the main parameter. Since in our case we have two parallel gradient boosting procedures with different learning rates, we reparameterize them as $(\lambda_{scale} , \lambda_{ratio}) = (\lambda^{\sigma}, \lambda^{\sigma}/\lambda^{\gamma})$. The balance described above is expressed between $B$ and $\lambda_{scale}$ and we propose the default $\lambda_{scale}=0.01$, leaving the number of trees $B$ as the primary parameter. The ratio of the learning rates is important as  $\gamma$ generally requires stronger regularization  than $\sigma$ and ranges on smaller scales. Therefore it is natural to choose $\lambda_{ratio} > 1$.  

\subsubsection{Remaining tuning parameters}
The minimum leaf sizes  $L_{\min}^\sigma, L_{\min}^\gamma$ and subsample fraction $s$ play the role of regularization parameters. The minimum leaf size makes sure that the splits do not try to isolate a single high observation of the gradient and that the leaves contain enough observations so that averaging provides a smoother gradient. Subsampling ensures that different trees are fitted on different sub-samples, mitigating the correlation between trees; see \cite{Friedman2002} and \cite[Section 10.12.2]{ESL} for further discussion on the regularization effect of subsampling. \tcb{It is common practice that early exploration determines suitable values for these parameters. Depending on the problem and the sample size, we recommend the range $[0.4, 0.8]$ for $s$ and $[10, \frac{n}{50}]$ for $L_{\min}$. }

The parameter $\tau_0$ stands for the probability level of the intermediate quantile used as threshold. Threshold selection is a long standing problem in extreme value theory \citep[e.g.,][]{dupuis1999exceedances, drees2020minimum}. A higher threshold yields a better approximation by the GPD distribution but fewer exceedances, leading to reduced bias and higher variance. Some guidelines for threshold selection in practice are provided in Section~\ref{sec:application}, where we present an application to precipitation forecast statistical post-processing.

\subsection{Tools for model interpretation}\label{sec:interpretation}
Contrary to a single tree, boosting models that aggregate hundreds or thousands of trees are difficult to represent but diagnostic plots are available to ease the interpretation. We briefly discuss variable importance and partial dependence plots, which are straightforward modifications to our framework of the tools detailed in \cite[Section 10.13]{ESL}.

\subsubsection{Variable importance} 
Boosting is quite robust to the curse of dimensionality and often provides good results even in the presence of high dimensional predictors and noise variables. Understanding which predictors are the most important is crucial for model interpretation. Variable importance is used for this purpose and we discuss here the permutation score and the relative importance. 

The permutation score helps to evaluate the impact of a predictor on the model deviance and is not specific to boosting. The relation between a predictor and the response is disturbed by shuffling the values of this predictor and measuring the difference in the deviance before and after shuffling. More precisely, for predictor $X_j$, we define 
\begin{equation}\label{eq: important_score}
I(X_j)=\sum_{i=1}^n\ell_{Z_i}\left(\hat \theta \left( \bb X^{(j)}_i  \right)\right)-\sum_{i=1}^n\ell_{Z_i}\left(\hat \theta \left( \bb X_i  \right)\right),     \end{equation}
where $\hat \theta$ is the estimator given in \eqref{eq:estimator} and  $\bb X^{(j)}_1,\ldots,\bb X^{(j)}_n$ denote the same input vectors as $\bb X_1,\ldots,\bb X_n$ except that  the $j$th components are randomly shuffled.  A large permutation score  $I(X_j)$ indicates a strong effect of $X_j$ in the boosting model. Since the scores are relative, it is customary to assign to the largest the value of $100$ and scale the others accordingly.

The relative importance is specific to tree based methods such as boosting or random forests and uses  the structure of the trees in the model. It is discussed for instance in \cite[Section 10.13.1]{ESL}. Recall that during the construction of the trees, the splits are performed so as to minimize the residual sum of squares (RSS) of the gradient and each split causes a decrease in the RSS. The more informative splits are those causing a large decrease in the RSS.
The relative importance of a given variable $X_j$ is obtained by considering all the splits due to this variable in the sequence of trees, and by summing up the decrease in RSS due to those splits. Because we have two sequences of trees, we compute relative importance of variable $X_j$ in the estimation of $\sigma$ and $\gamma$ separately by considering the sequence of trees $(T_b^\sigma)$ and $(T_b^\gamma)$ respectively.

\subsubsection{Partial dependence plot} 
Once the most relevant variables have been identified, the next attempt is to understand the dependence between the predictors and the response. Partial dependence plots  offer a nice graphical diagnostic of the partial influence of a predictor $X_j$ on the outputs $\hat\sigma(\bb x)$, $\hat\gamma(\bb x)$ or $\hat Q_{\bb x}(\tau)$; see \cite[Section 10.13.2]{ESL}. The partial dependence plot for $\hat\sigma$ with respect to $X_j$ is the graph of the function
$x\mapsto \frac{1}{n}\sum_{i=1}^n \hat\sigma(\bb X^{-j,x}_i)$, where the vector  $\bb X^{-j,x}_i$ is equal to $\bb X_i$ except that the $j$th component has been  replaced by $x$. Notice that dependence between the predictors is not taken into account so that this is not an estimate of $\mathbb{E}[\hat\sigma(\bb X)\mid X_j=x]$, except if $X_j$ is independent of the other predictors. In the particular case when an additive model is built, i.e., $D^\sigma=1$, the partial dependence plot with respect to $X_j$ is equal to the effect of the variable $X_j$ up to an additive constant. Partial dependence plots with respect to several covariates can  be defined and plotted similarly, at least in dimension $2$ or $3$.

\section{Simulation studies}\label{sec:num-exp}
To demonstrate the performance of our method, we conduct two numerical experiments. We generate $n$ independent samples with $d$ covariates $\bb X=(X_1,\ldots, X_{d})$  distributed from an independent uniform distribution on $[-1,1]^d$,  with $(n, d)=(2000, 40)$ or $(5000, 10)$, depending on the complexity of the model.  We aim to estimate the conditional quantile function $Q_{\bb x}(\tau)$ corresponding to extreme probability levels $\tau\in\{0.99, 0.995,0.9995\}$.  We choose the level $\tau_0=0.8$ for the intermediate quantile and it is worthwhile to note that the effective sample size $n(1-\tau_0)$ for the gradient boosting step is then only $400$ for $n=2000$.   

The local smoothing based methods mentioned in the introduction \citep{GardesStupfler2019, Daouia2013} become cumbersome in our simulation setting because of the sparsity of data in high dimension.  We compare our \texttt{gbex} method to two quantile regression approaches, the quantile regression forest (\texttt{qrf}) from \cite{Meinshausen2006} and the generalized random forest (\texttt{grf}) from \cite{Athey2019}. Moreover, we consider two existing methods from extreme value theory that use GPD modeling of the exceedances. One is the classical estimator of extreme quantile without using covariates, thus $\gamma(\bb x)\equiv \gamma$ and $\sigma(\bb x)\equiv \sigma$, which we call the \texttt{constant} method. The other one is the \texttt{evgam} method of \cite{you2019} that assumes generalized additive models for $\gamma(\bb x)$ and $\sigma(\bb x)$. 

To evaluate the performance over the full predictor domain $[-1,1]^d$ we consider the integrated squared error (ISE) defined for a fixed quantile level $\tau$ and the $i$th replication of the data set by 
\begin{equation}
{\rm ISE}_i=\int_{[-1, 1]^d} \left(  \hat Q^{(i)}_{\bb x}(\tau )-Q_{\bb x}(\tau ) \right)^2 d\bb x, \label{eq: ISE}
\end{equation}
where $\hat Q^{(i)}_{\bb x}(\tau )$ is the quantile estimated from the model. We use a Halton sequence, a low discrepancy quasi-random sequence \citep[e.g.,][p.~29]{Niederreiter1992}, in order to efficiently evaluate the high dimensional integral in the ISE computation. Averaging over the $R = 1000$ replications, we obtain the mean integrated squared error (MISE).

Our first model is designed to check robustness of the methods against noise variables. This model is constructed in a similar way as the example studied in \cite[][Section 5]{Athey2019} and it has a predictor dimension of $d=40$, of which one covariate is signal and the remaining are noise variables. 
\begin{itemize}
\item {\bf Model 1:} Given $\bb X = \bb x \in \RR^{40}$, $Y$ follows a Student's $t$-distribution with $4$ degrees of freedom and scale $$\mathrm{scale}(\bb x)=1+\mathds{1}(x_1>0).$$ 

This is a heavy-tailed model where the  GPD approximation has a constant shape parameter $\gamma(\bb x)\equiv 1/4 $ and the scale parameter is a step function in $X_1$. More precisely, $\sigma(\bb x)=\sigma(\tau_0)(1+\mathds{1}(x_1>0))$ where $\sigma(\tau_0)$ is a multiplicative constant depending on the threshold parameter $\tau_0$.
\end{itemize}

In our second model, we consider a more complex response surface where  both the scale and shape parameters depend on the covariates and interactions of order $2$ are introduced.
\begin{itemize} 
\item {\bf Model 2:}  Given $\bb X=\bb x \in \RR^{10}$, $Y$ follows a Student's $t$-distribution with degree of freedom ${\rm df}(\bb x)$ depending on $x_1$ through
 $${\rm df}(\bb x)=7\left(1+\exp(4x_1+1.2)\right)^{-1}+3,$$ 
and scale parameter $\mathrm{scale}(\bb x)$ depending on $(x_1,x_2)$ through
$$\mathrm{scale}(\bb x)=1+6\varphi(x_1, x_2),$$
where $\varphi$ denotes the  density function of a bivariate normal distribution with standard normal margins and correlation $0.9$. The numerical constants are chosen so that the GPD approximation of $Y$ given $\bb X=\bb x$ has parameters $\gamma(\bb x)= 1/{\rm df}(\bb x)$  in the range $[0.10, 0.33]$ for $\bb x \in[-1,1]^d$, $d=10$.
\end{itemize}

\subsection{Tuning parameters and cross validation}
We generate samples of size $n=2000$ and 5000, respectively from Model~1 and Model~2. We set the following tuning parameters for \texttt{gbex}: the learning rate $\lambda_{scale}=0.01$ and  the sample fraction $s=75\%$ for both models; $\lambda_{ratio}=15$ for Model~1 and $\lambda_{ratio}=7$ for Model~2.  

\begin{figure}[hbt!]
    \centering
    \begin{minipage}{.5\textwidth}
        \centering
        \includegraphics[width=0.9\linewidth]{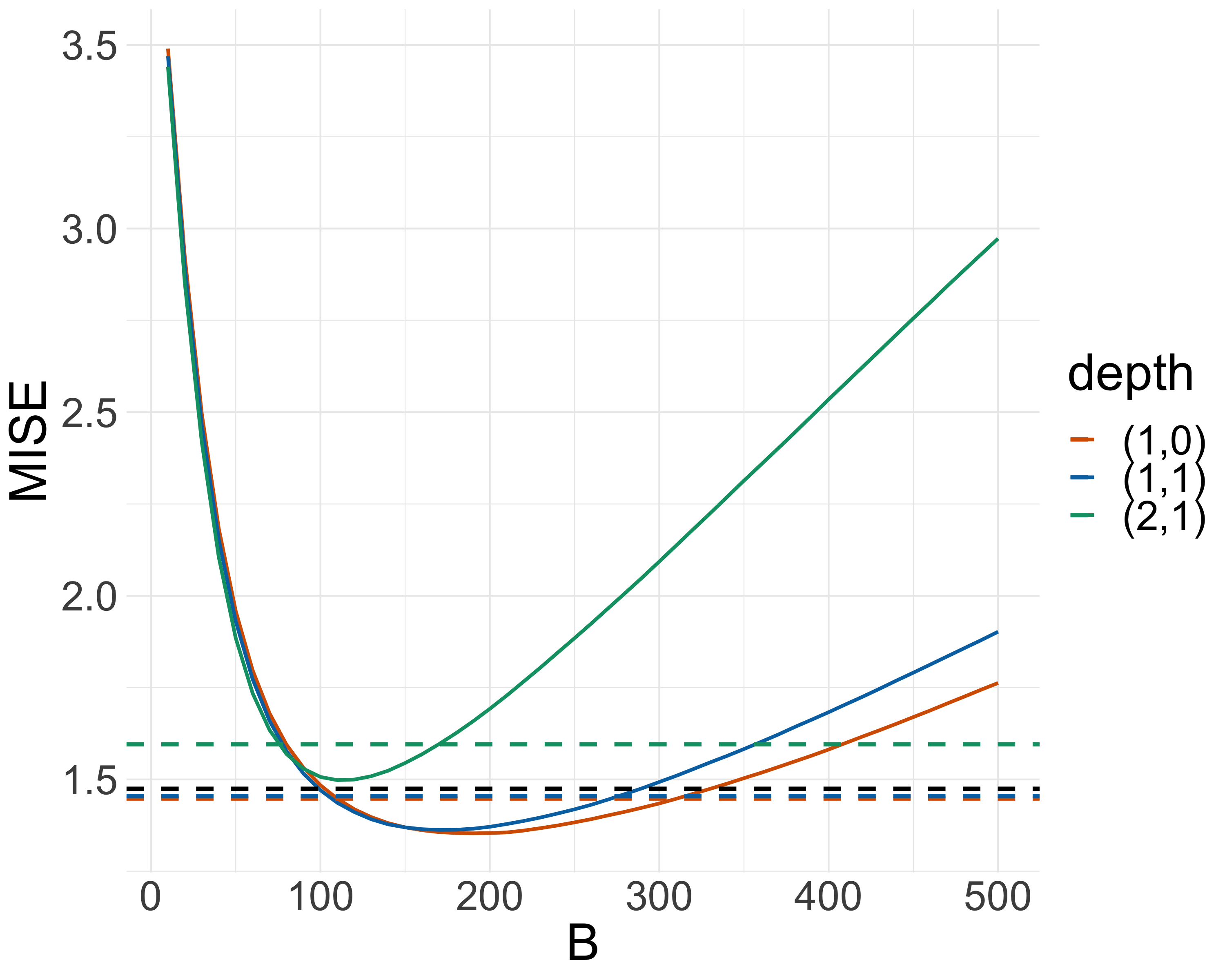}
  
    \end{minipage}%
    \begin{minipage}{0.5\textwidth}
        \centering
        \includegraphics[width=0.9\linewidth]{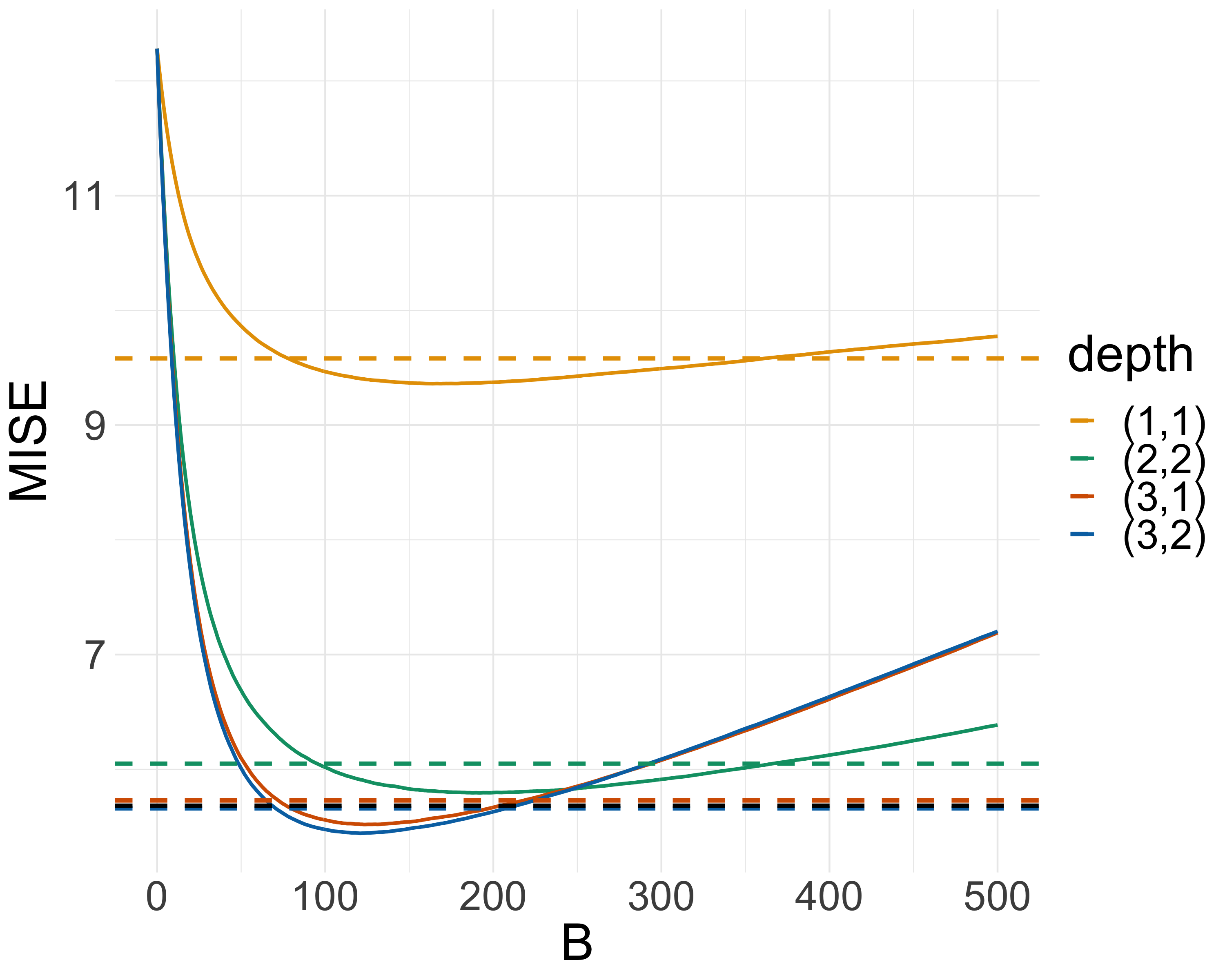}
    \end{minipage}
\caption{The MISE for Model 1 (left panel) and Model 2 (right panel)  of the \texttt{gbex} extreme quantile estimator with probability level $\tau=0.995$ as a function of $B$ for various depth parameters (curves); the MISE of the \texttt{gbex} estimator with adaptive choice of $B$  for various depth parameters (horizontal dashed lines); the MISE of the \texttt{gbex} estimator with both tree number and depth parameters selected by cross-validation (black dashed line). }
    \label{fig: B_depth_CV}
\end{figure}

As discussed in Section \ref{sec:tuning}, the number of trees $B$ is the most important regularization parameter and the depth parameters $(D^\sigma, D^\gamma)$ determine  the complexity of the fitted model. Therefore, we investigate how these tuning parameters influence the performance of our estimator in terms of MISE. Figure \ref{fig: B_depth_CV} shows the results for Model~1 (left panel) and for Model 2 (right panel). The curves  represent the  MISE of \text{gbex} as a function of $B$ for various depth parameters  $(D^\sigma, D^\gamma)$.  The right panel clearly shows that for Model 2 the choice $(D^\sigma, D^\gamma)=(1, 1)$ does not account for the model complexity adequately, which leads to a high MISE. Indeed, boosting with depth one tries to fit an additive model but the scale parameter of Model 2 depends on $(X_1,X_2)$  in a non-additive way. On the other hand, for Model 1, which is an additive model with the optimal depth  $(D^\sigma, D^\gamma)=(1, 0)$,  the curves  suggest that assuming unnecessary complexity of the model might lead to suboptimal behavior of the estimator: the choice $(2, 1)$ yields higher MISE  than the other two choices and the MISE stays low for a shorter range of $B$. In general, higher depths help the model to adapt the data faster but then overfitting is prone to occur more rapidly when $B$ increases. 
The horizontal dashed lines in  Figure \ref{fig: B_depth_CV} represent the resulting MISE of our estimator when $B$ is chosen  via cross-validation with deviance loss given in \eqref{eq: CV_D}, with $K=5$ folds and $10$ replications. The plots confirm that  the data driven choice of $B$ results in near optimal MISE  for fixed depth parameters (with dashed horizontal lines close to the minimum of the curve with the same color). We additionally apply cross-validation to select both $B$ and $(D^\sigma, D^\gamma)$ simultaneously. The resulting MISE is represented by the black dashed line, which is very close to the minimum of all the dashed lines. Overall, the results confirm the good performance of the proposed cross-validation procedure. 

For the rest of the simulation study, we set  $(D^\sigma, D^\gamma)=(1, 1)$ for Model 1 and $(D^\sigma, D^\gamma)=(3, 1)$ for Model 2 and choose $B$ with  cross-validation.

\subsection{Comparison with different methods}
\begin{figure}[hbt!]
\begin{center}
\includegraphics[width=0.9\textwidth]{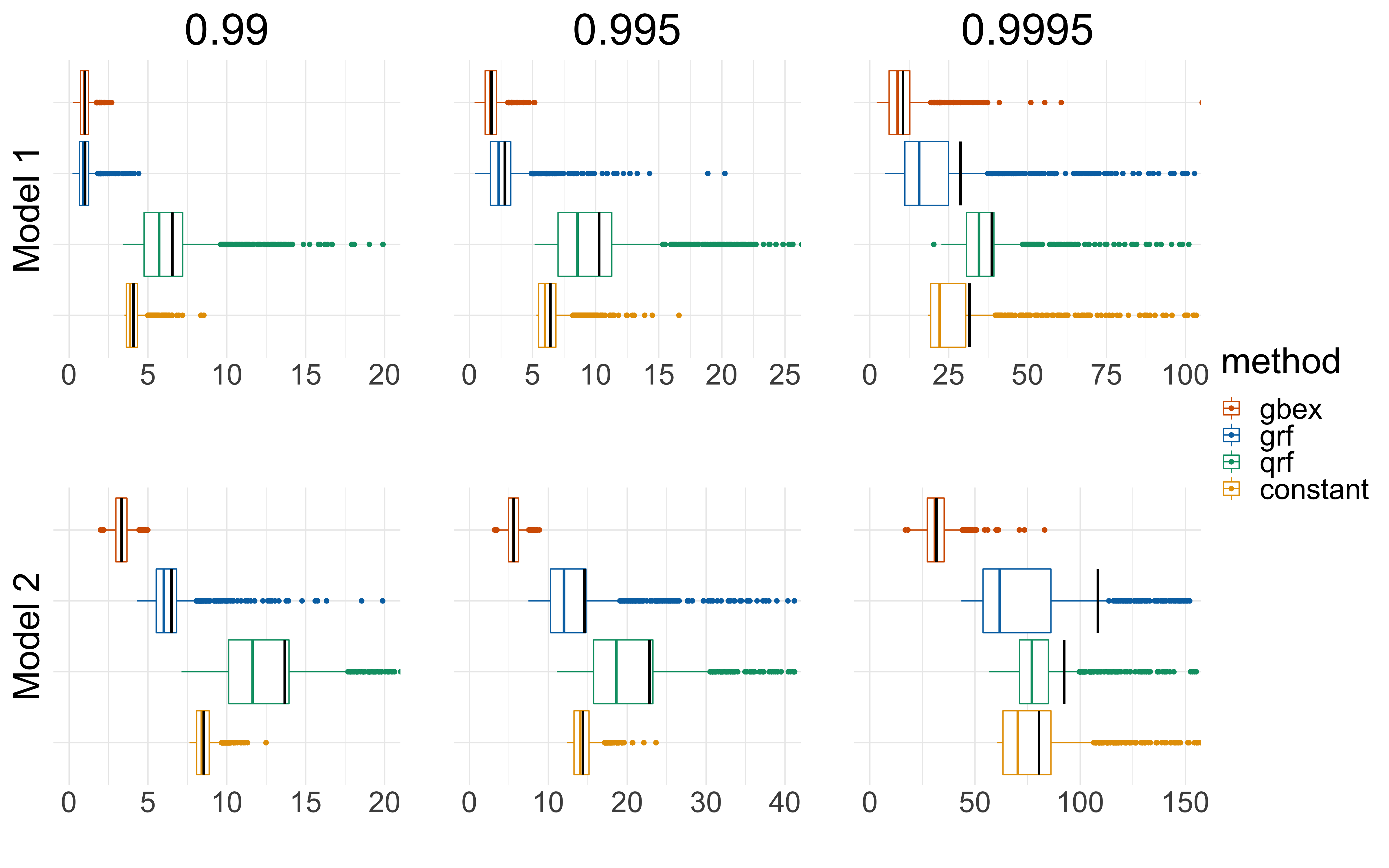}
\caption{Boxplot of ISE based on 1000 replications for the four quantile estimators (\texttt{gbex}, \texttt{grf}, \texttt{qrf} and \texttt{constant}) at different probability levels $\tau=0.99$ (left), $0.995$ (middle) and $0.9995$ (right) for Model~1 (top) and Model~2 (bottom).  Some outliers of  \texttt{grf} and  \texttt{qrf} are left out for a clearer comparison.  The black vertical lines indicate the MISE.}
\label{fig: MISE-c}
\end{center}
\end{figure}

The comparison of our \texttt{gbex} method to the other three approaches \texttt{qrf}, \texttt{grf} and \texttt{constant}, is presented in Figure~\ref{fig: MISE-c}. The  results for Model~1 and Model~2 are given in the first and second row, respectively. For the probability level $\tau=0.99$, $0.99$ and $0.9995$ in the left, middle and right column, the figure shows the boxplots of ISE defined in \eqref{eq: ISE} and the MISE represented by the vertical black line. The MISE  grows as the probability level increases for all methods, however \texttt{gbex} clearly outperforms the other three approaches with a much smaller MISE and a much lower variation of ISE. When the probability level $\tau$ is close to or larger than $(1-1/n)$ (right column), both \texttt{grf} and \texttt{qrf} lead to  extremely large ISE outliers so that the ISE mean is larger than the  third quartile (black line outside the box). Some extreme outliers of ISE are left out of the boxplots to have a clear comparison.  \tcb{We have also investigated other models (Burr, GPD) and the comparison results are reported in Appendix \ref{app: add sim}.  }

In Figure~\ref{fig: MISE-c}, we have not included \texttt{evgam}, the main competitor from extreme value theory. The reason is that for high-dimensional predictor spaces with many noise variables as in our simulations, the additive model for the GPD parameters suffers severely from the curse of dimensionality. Indeed, in an additional simulation from Model~1 with a varying dimension $d$ of the predictor space (not shown here), 
the MISE of \texttt{evgam} grows quickly as a function of $d$. The MISE of \texttt{gbex}, on the other hand, remains fairly constant with growing number of noise variables. The simulation result reveals that the MISE of \texttt{gbex}  with $d=40$ is similar to the MISE of \texttt{evgam} with $d=4$.
This underlines the robustness of \texttt{gbex} against the curse of dimensionality and noise variables, which is a prominent advantage of tree based methods.

\subsection{Diagnostic plots}
\begin{figure}[hbt!]
    \centering
    \begin{minipage}{.5\textwidth}
        \centering
        \includegraphics[width=0.9\linewidth]{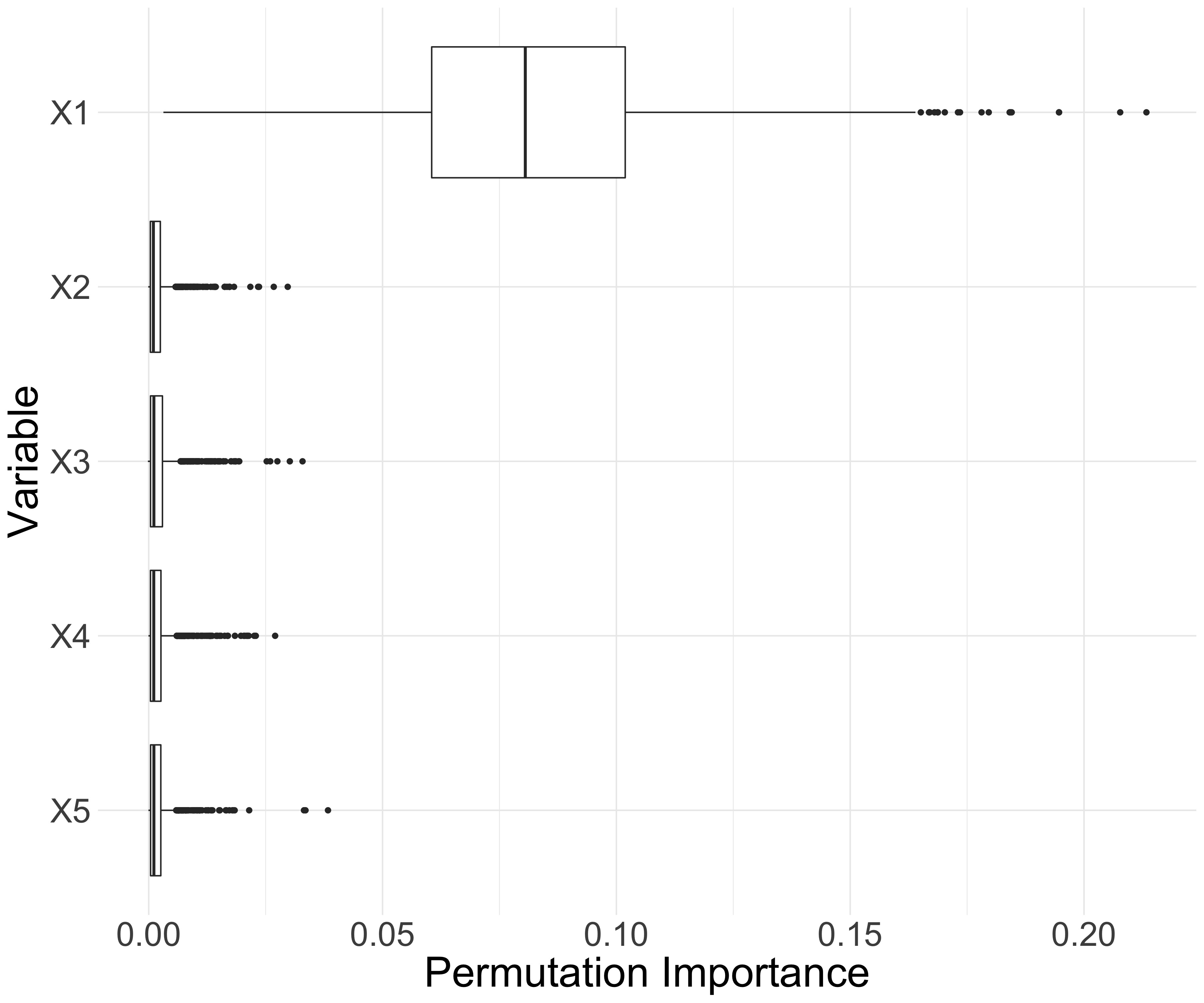}
  
    \end{minipage}%
    \begin{minipage}{0.5\textwidth}
        \centering
     \includegraphics[width=0.9\linewidth]{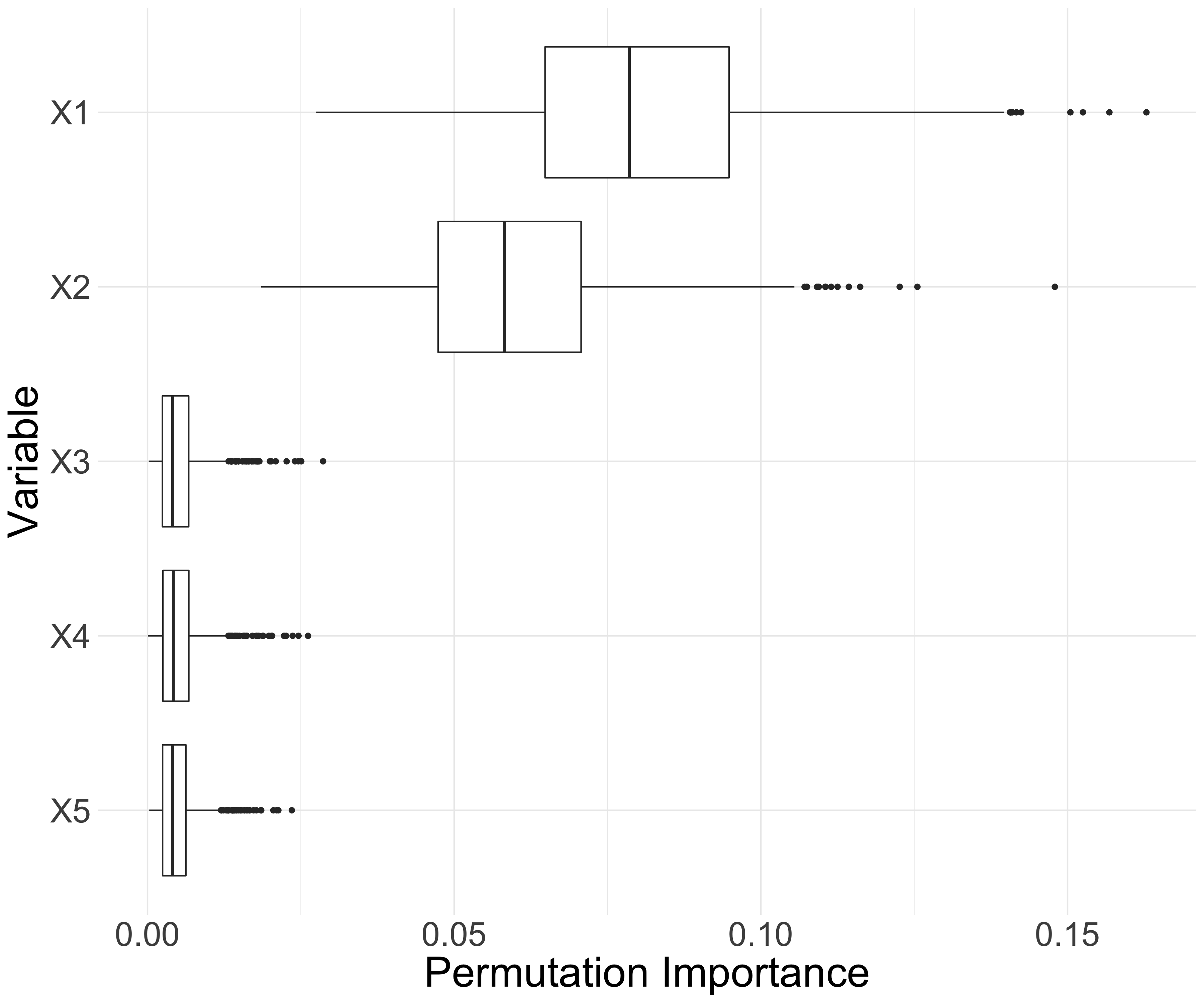}
    \end{minipage}
\caption{Boxplots of permutation  scores defined in \eqref{eq: important_score} for $X_j$, $j=1,\ldots, 5$, based on 1000 samples. Left panel: Model 1, where only $X_1$ contains signal. Right panel:  Model 2, where only $X_1$ and $X_2$ contain signal.}
\label{fig: Var-score}
\end{figure}

We finally look at the model interpretation diagnostics. Figure \ref{fig: Var-score} shows the permutation importance scores defined in \eqref{eq: important_score}  for both models, based on 1000 replications.  The boxplots clearly show that this score is able to  identify the signal variable(s).  Note that there are 39 noise variables for Model 1 and 8 for Model 2. The scores of the noise variables behave all similarly and only a limited number are displayed.  For Model 2, the permutation score is higher for $X_1$ than for $X_2$, due to the fact that $X_1$ contributes to both shape and scale functions while $X_2$ only contributes to the scale function.

\begin{figure}[hbt!]
    \centering
    \begin{minipage}{.5\textwidth}
        \centering
        \includegraphics[width=0.9\linewidth]{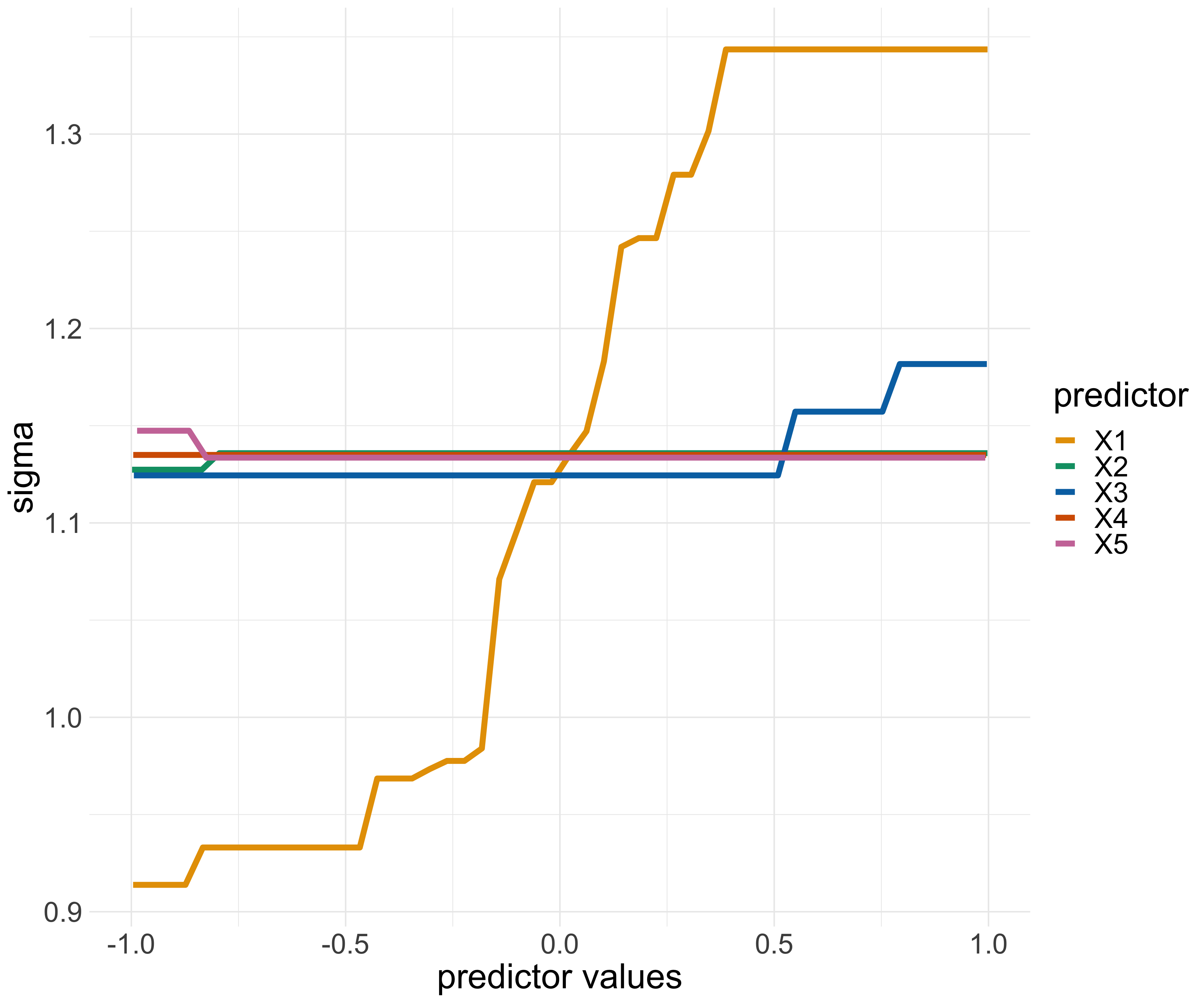}
  
    \end{minipage}%
    \begin{minipage}{0.5\textwidth}
        \centering
     \includegraphics[width=0.9\linewidth]{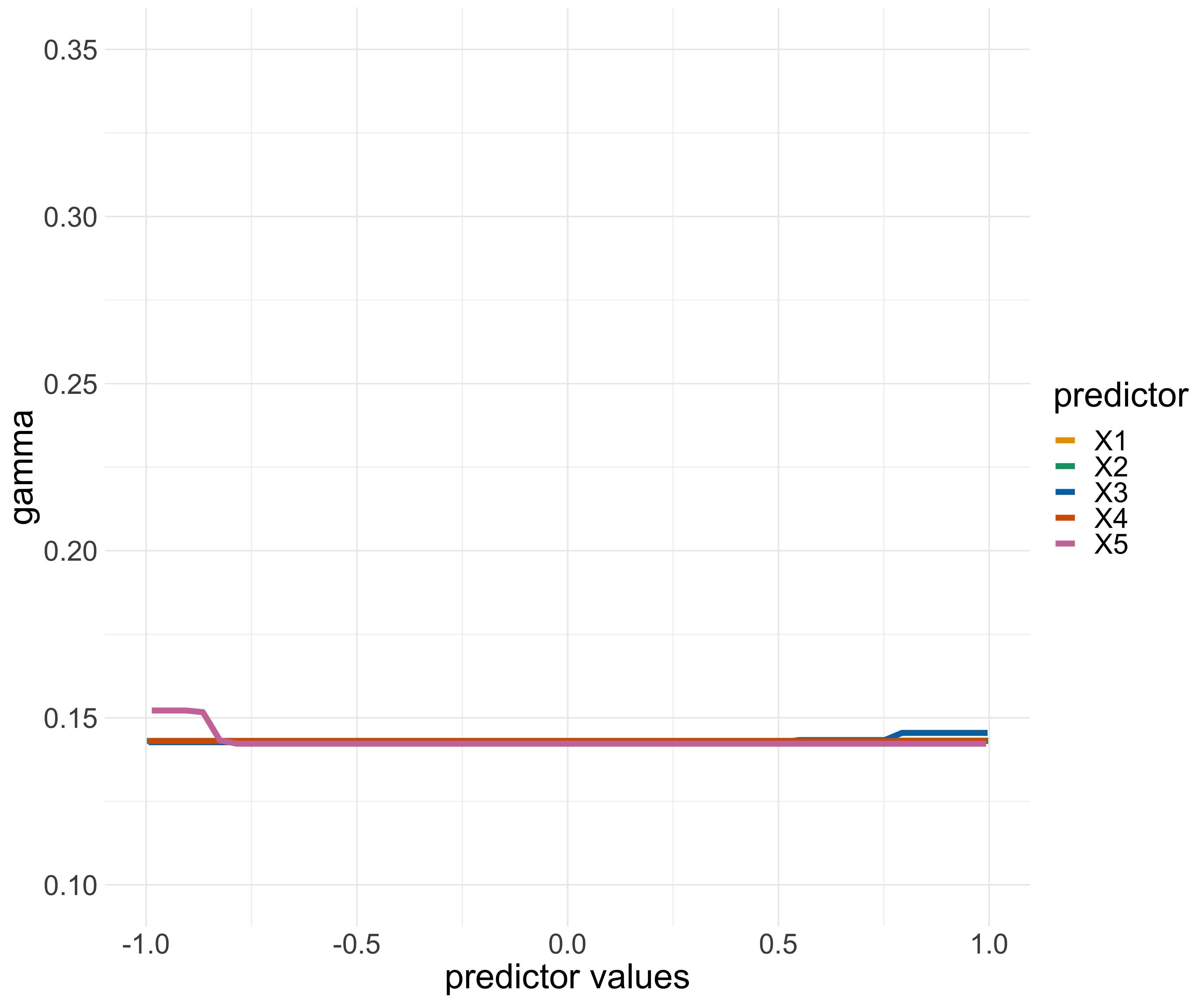}
    \end{minipage}
    \caption{Partial dependence plots of $\hat\sigma$ (left panel) and of $\hat \gamma$ (right panel) with respect to $X_j$, $j=1,\ldots, 5$, based on one random sample of Model 1. }
    \label{fig: model_interpretation}
\end{figure}

\begin{figure}[H]
    \centering
    \begin{minipage}{.5\textwidth}
        \centering
        \includegraphics[width=0.9\linewidth]{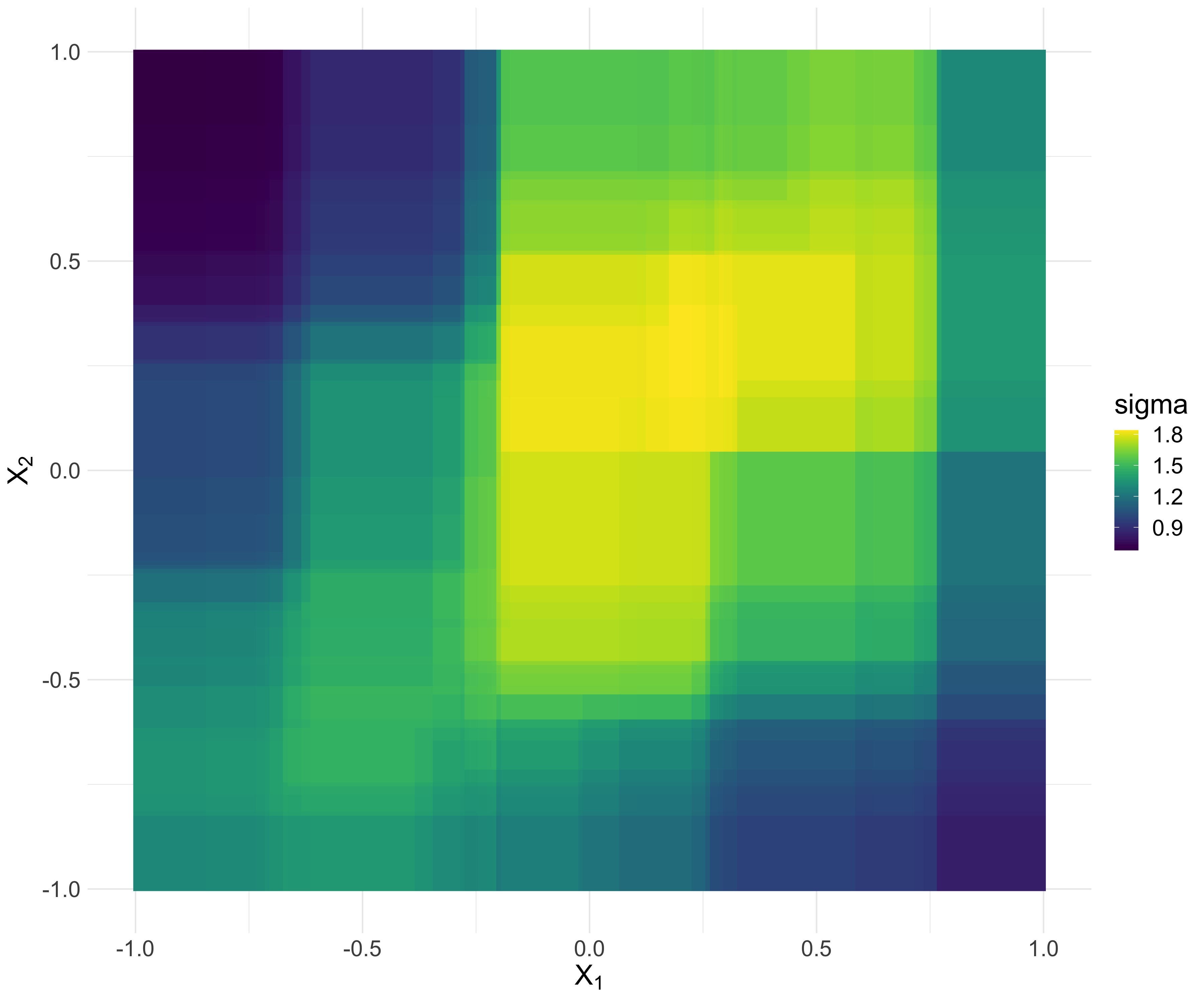}
  
    \end{minipage}%
    \begin{minipage}{0.5\textwidth}
        \centering
     \includegraphics[width=0.9\linewidth]{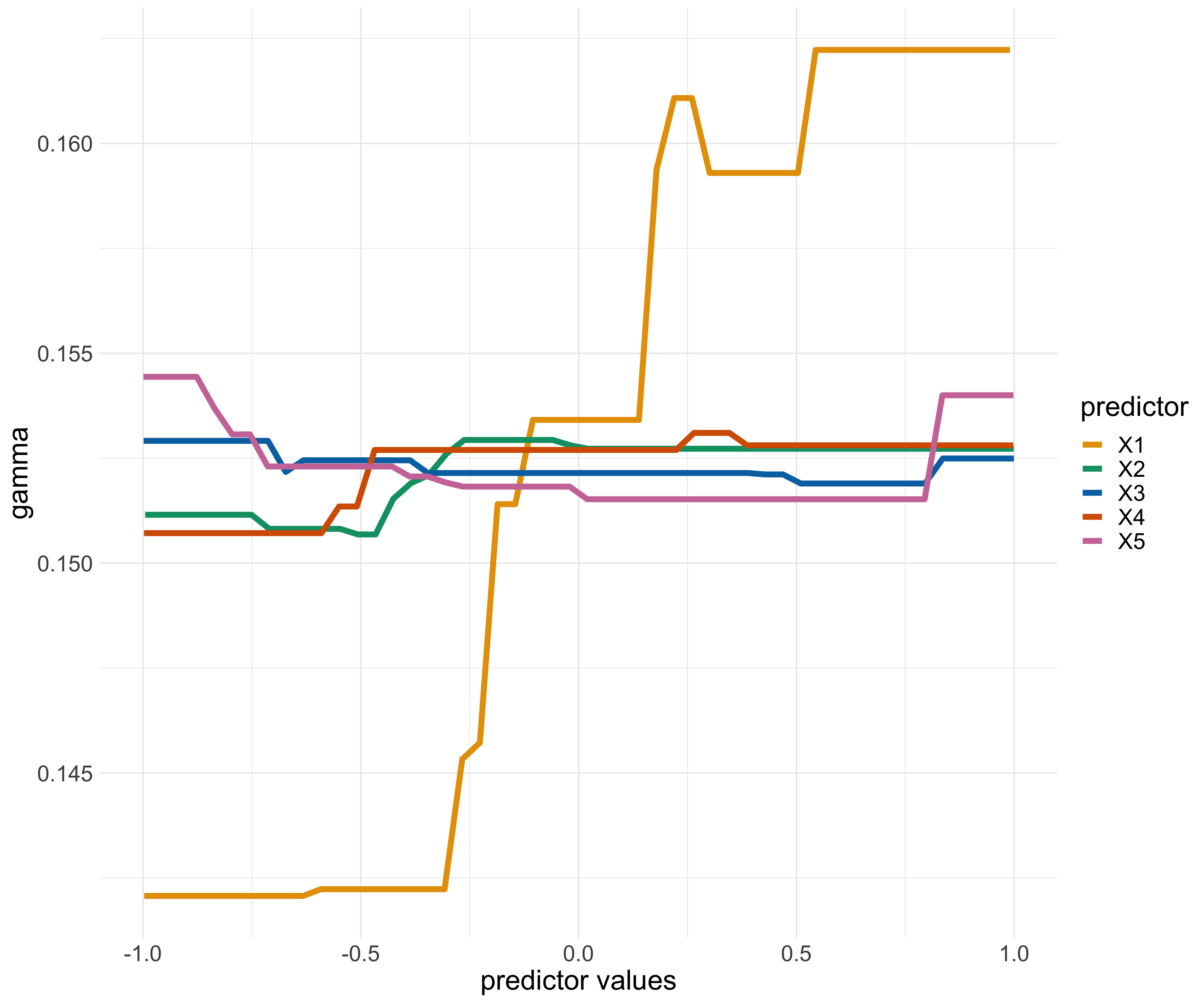}
    \end{minipage}
    \caption{Left panel: partial dependence plots of $\hat\sigma$ with respect to $(X_1,X_2)$. Right panel: partial dependence plot of $\hat \gamma$ with respect to $X_j$, $j=1,\ldots, 5$. Both experiments corresponds to one random sample of Model 2.}
        \label{fig: PD_model 2}
\end{figure}

The left panel of Figure \ref{fig: model_interpretation} presents a typical partial dependence plot (Section \ref{sec:interpretation}) for $\hat\sigma$ based on one random sample from Model 1. This plot clearly suggests that $\hat \sigma$ is  a step function of $X_1$ and does not depend on the noise variables.  The partial dependence plot for $\hat \gamma$ indicates that the shape does not change with respect to any of the covariates. For this model, the partial dependence plots are in perfect agreement with the simulation design. For Model 2, the left panel of Figure \ref{fig: PD_model 2} shows the partial dependence plot of the scale parameter with respect to $X_1$ and $X_2$. We see that the model detects the right pattern of larger values on the diagonal and in the center. The right panel shows that the model identifies the impact of $X_1$ on the shape parameter while the partial dependence plot of the other variables is fairly constant, again in agreement with the simulation design.

\section{Application to precipitation forecast}\label{sec:application}
Extreme precipitation events can have disruptive consequences on our society. Accurate predictions are vital for taking preventive measures such as pumping water out of the system to prevent flooding. We apply our \texttt{gbex} method to predict extreme quantiles of daily precipitation using the output of numerical weather prediction (NWP) models. 

Weather forecasts rely on NWP models that are based on non-linear differential equations from physics describing the atmospheric flow. The solutions to these equations  with respect to  initial conditions and parametrizations of  unresolved processes form a forecast that is deterministic in nature. Introducing uncertainty in these initializations yields an ensemble forecast that consists of multiple members. In this application, we use the ensemble forecast from the European Centre for Medium-Range Weather Forecasts (ECMWF) as covariates in \texttt{gbex} to predict the daily precipitation. Using NWP output for further statistical inference to improve forecasts is known as statistical post-processing. 

\subsection{Precipitation Data}
Our data set consists of ECMWF ensemble forecasts of daily accumulated precipitation and the corresponding observations at seven meteorological stations spread across the Netherlands (De Bilt, De Kooy, Eelde,   Schiphol,  Maastricht, Twente and Vlissingen)\footnote{Observed daily precipitation can be obtained from \url{http://projects.knmi.nl/klimatologie/daggegevens/selectie.cgi}}. We use about 9 years of data, from January 1st, 2011, until November 30th, 2019, with sample size $n=3256$.  We fit separate models for each station with response variable $Y$ equal to the observed precipitation at the station between 00 UTC and 24 UTC. 

As for the covariates, we use ECMWF ensemble forecasts of daily accumulated precipitation that is computed the day before at 12UTC. 
The  ensemble forecast contains $51$  members. For efficiency, we use  two summary statistics, namely the standard deviation of the ensemble members and the upper order statistics (the maximum of the ensemble members). Because most part of the Netherlands is flat and the distance between stations is not large, we include the ensemble summary statistics of all stations as covariates for the model of each station. To account for seasonality, we additionally consider the sine and cosine with a period of 365 for the day of the year. The total covariate dimension   is $d = 7\times 2 +2=16$,  for each model. We denote our data as $(Y_i^{(l)}, \mathbf{X}_i)$, where $\mathbf{X}_i\in\mathbb{R}^{16}$, $i=1,\ldots, n=3256$ and $l=1,\ldots, 7$. For station $l$, we apply the \texttt{gbex} Algorithm~\ref{algo:gbex2} to  $\{(Y_i^{(l)}, \mathbf{X}_i), i=1,\ldots, n\}$ to obtain  estimates of $Q^{(l)}_{\mathbf{X}}(\tau)$.

\subsection{Model fitting}
For model fitting, we have observed in a preliminary analysis that the output is sensitive to the initial value of $(\gamma,\sigma)$ and we propose a specific strategy that provides better results than the default initialization. We consider a common initial value for the shape $\gamma$ for all the stations and different initial values of $\sigma$ for the different stations, which leads to $\theta_0=(\gamma,\sigma_1,\ldots,\sigma_7)$. More precisely, we obtain the initial values by optimizing the log-likelihood function
\begin{equation*}\label{eq:param-init-data}
	L(\theta_0) = \sum_{l=1}^7 \sum_{i=1}^n \left[(1+1/\gamma)\log\left(1+\gamma \frac{Y_i^{(l)} - c}{\sigma_l}\right) + \log \sigma_l\right]\mathds{1}_{\{Y_i^{(l)}- c>0\}},
\end{equation*}
where $c$ is a large threshold chosen such that the estimate of $\gamma$ becomes stable.

\begin{figure}[hbt!]
    \centering
    \begin{minipage}{.5\textwidth}
        \centering
        \includegraphics[width=0.95\linewidth]{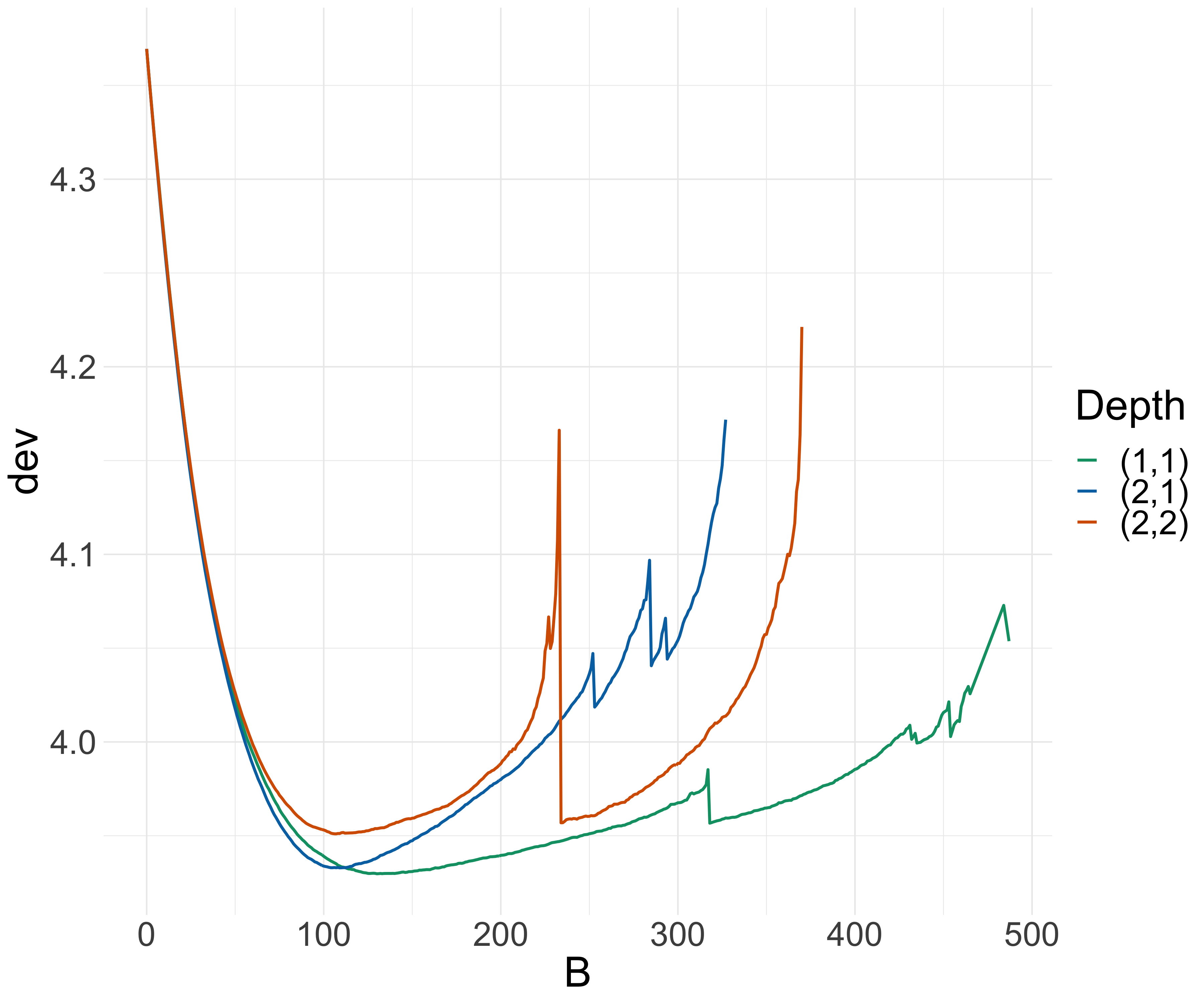}
  
    \end{minipage}%
    \begin{minipage}{0.5\textwidth}
        \centering
     \includegraphics[width=0.95\linewidth]{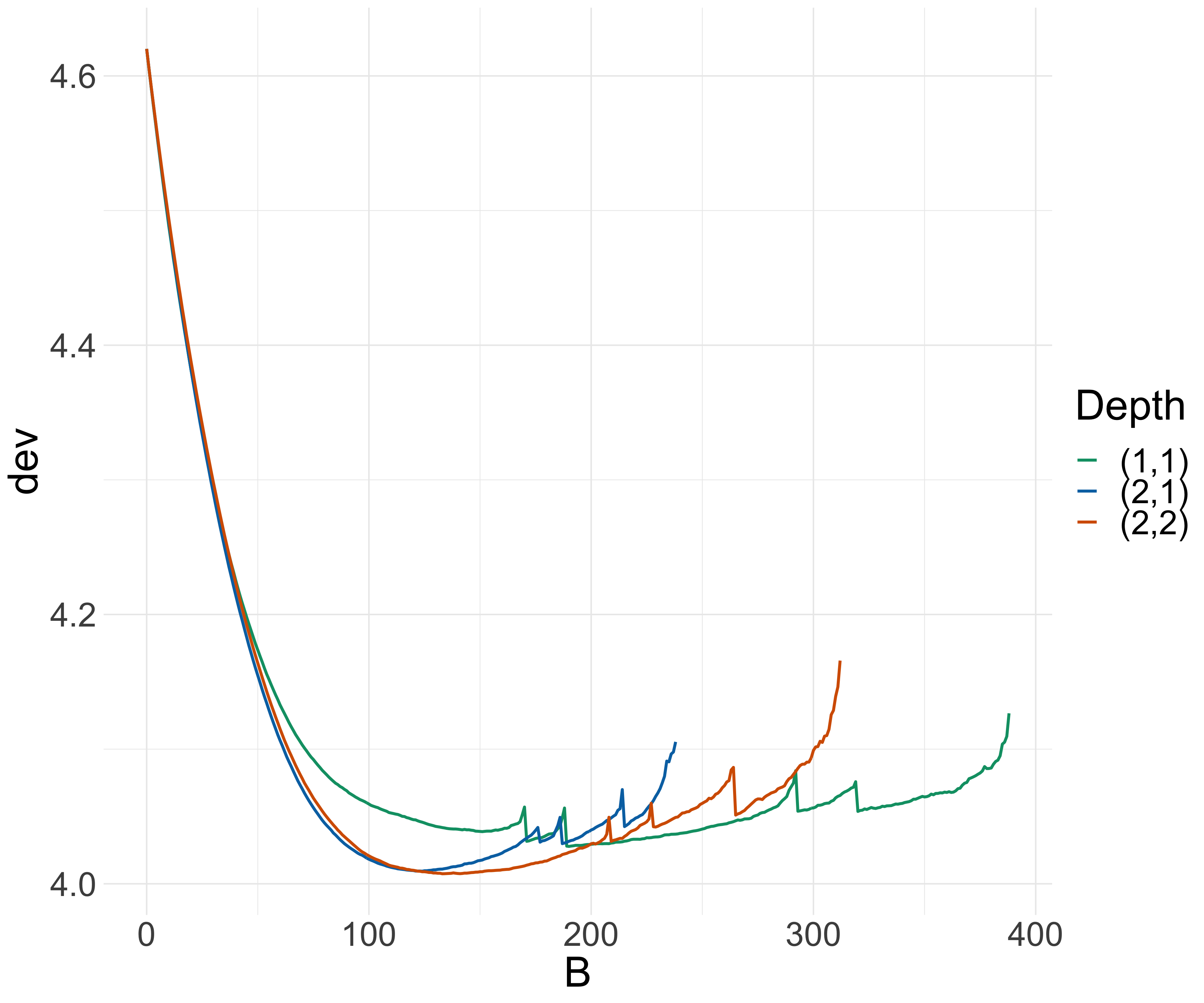}
    \end{minipage}    
    \caption{Cross-validation deviance given by \eqref{eq: CV_D} against $B$ for the data at stations Eelde (left) and Schiphol (right) in the application in Section~\ref{sec:application}.}   
    \label{fig:CV_app}
  \end{figure}

We apply \texttt{gbex} as detailed in Algorithm~\ref{algo:gbex2} with $\tau_0=0.8$ for each model.  We choose all tuning parameters except for $B$ to be the same for the seven models, in such a way to achieve the overall best combined deviance score for all stations. This prevents overfitting for a specific station and it results in the following choices:
$
(D^{\sigma}, D^{\gamma}) = (2,1),  (\lambda_{scale},\lambda_{ratio}) = (0.01,12), s=50\%,
\text{and } (L^{\sigma}_{\min}, L^{\gamma}_{\min}) = (15, 45).
$
Figure \ref{fig:CV_app} shows the cross-validated deviance as a function of the number of trees $B$ for different depth levels at two stations. The deviance behaves quite similar for the two stations and we choose $(D^{\sigma}, D^{\gamma}) = (2,1)$ for all stations. The optimal $B$ for each station is then chosen as the minimizer of the cross-validated deviance.

\subsection{Results}
\begin{figure}[hbt!]
       \begin{minipage}{.5\textwidth}
        \centering
        \includegraphics[width=0.9\linewidth]{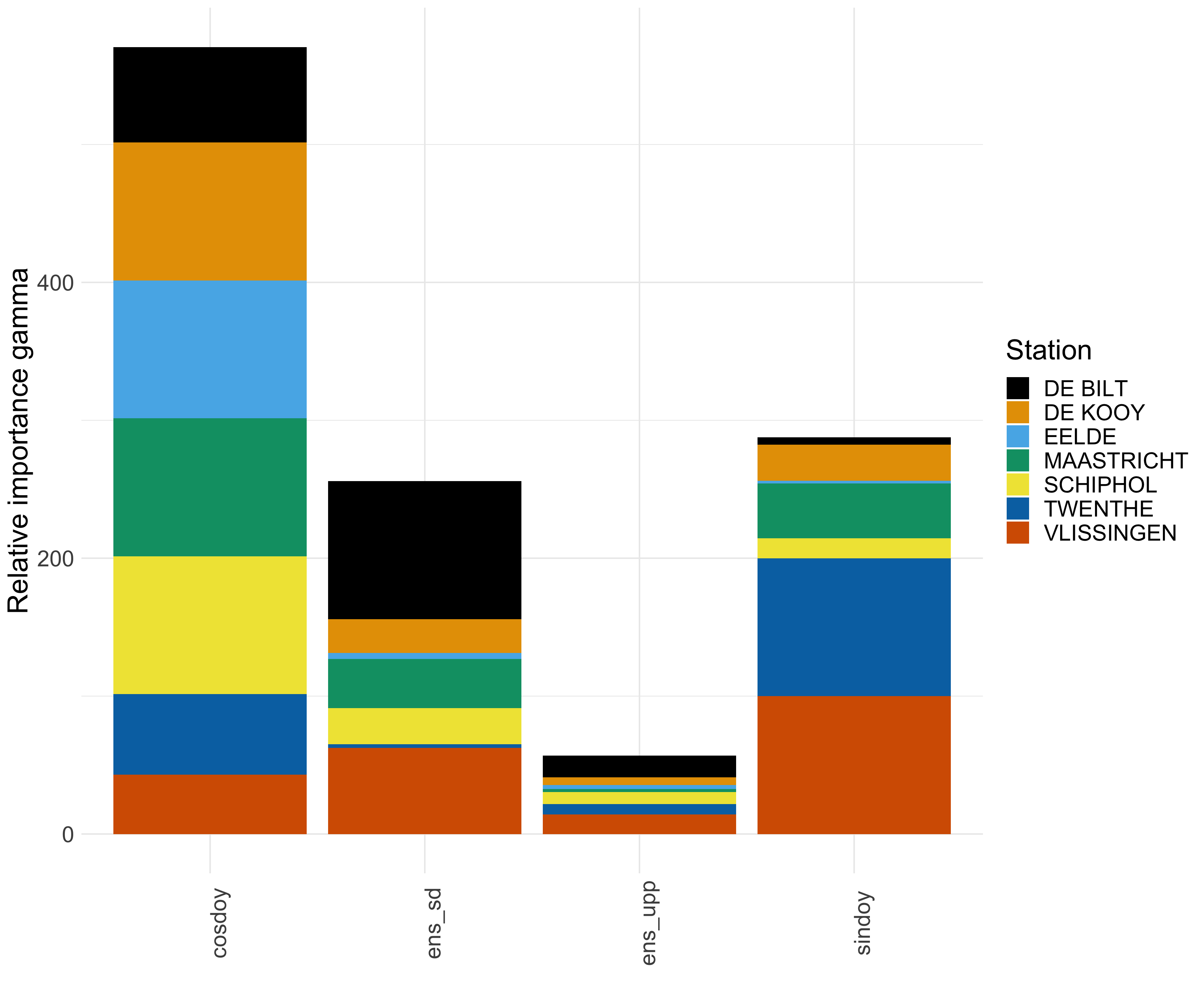}
    \end{minipage}%
    \begin{minipage}{0.5\textwidth}
        \centering
     \includegraphics[width=0.9\linewidth]{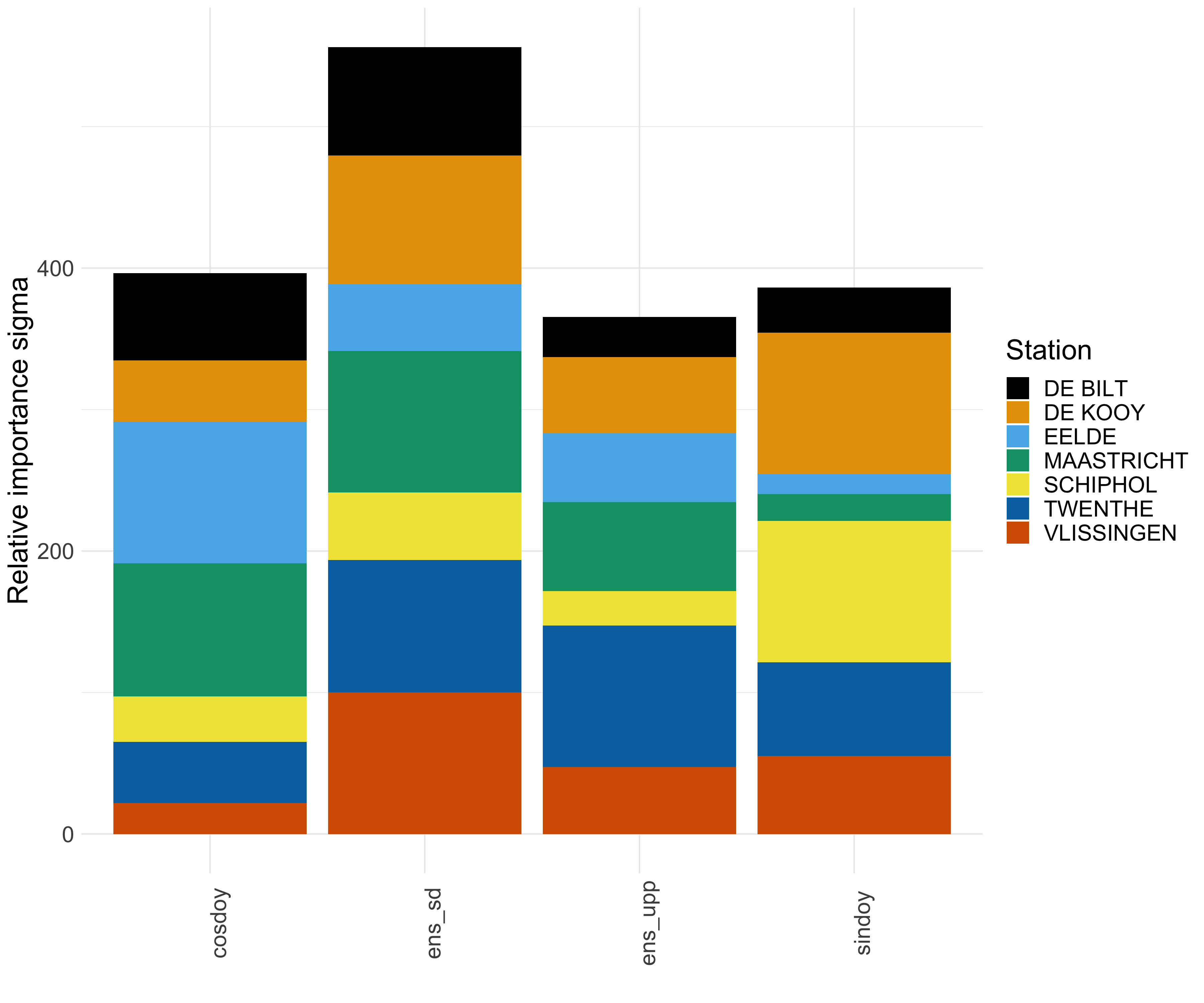}
    \end{minipage}
       \caption{Relative variable importance score for $\gamma$ (left) and $\sigma$ (right). For each model, the scores are normalized such that the maximum score is 100. \tcb{The scores for the variables \texttt{ens\_sd} (resp. \texttt{ens\_up}) at the 7 stations are aggregated into a single score}.}
    \label{fig:VI_relative}
\end{figure}

\begin{figure}[hbt!]
       \begin{minipage}{.5\textwidth}
        \centering
        \includegraphics[width=0.9\linewidth]{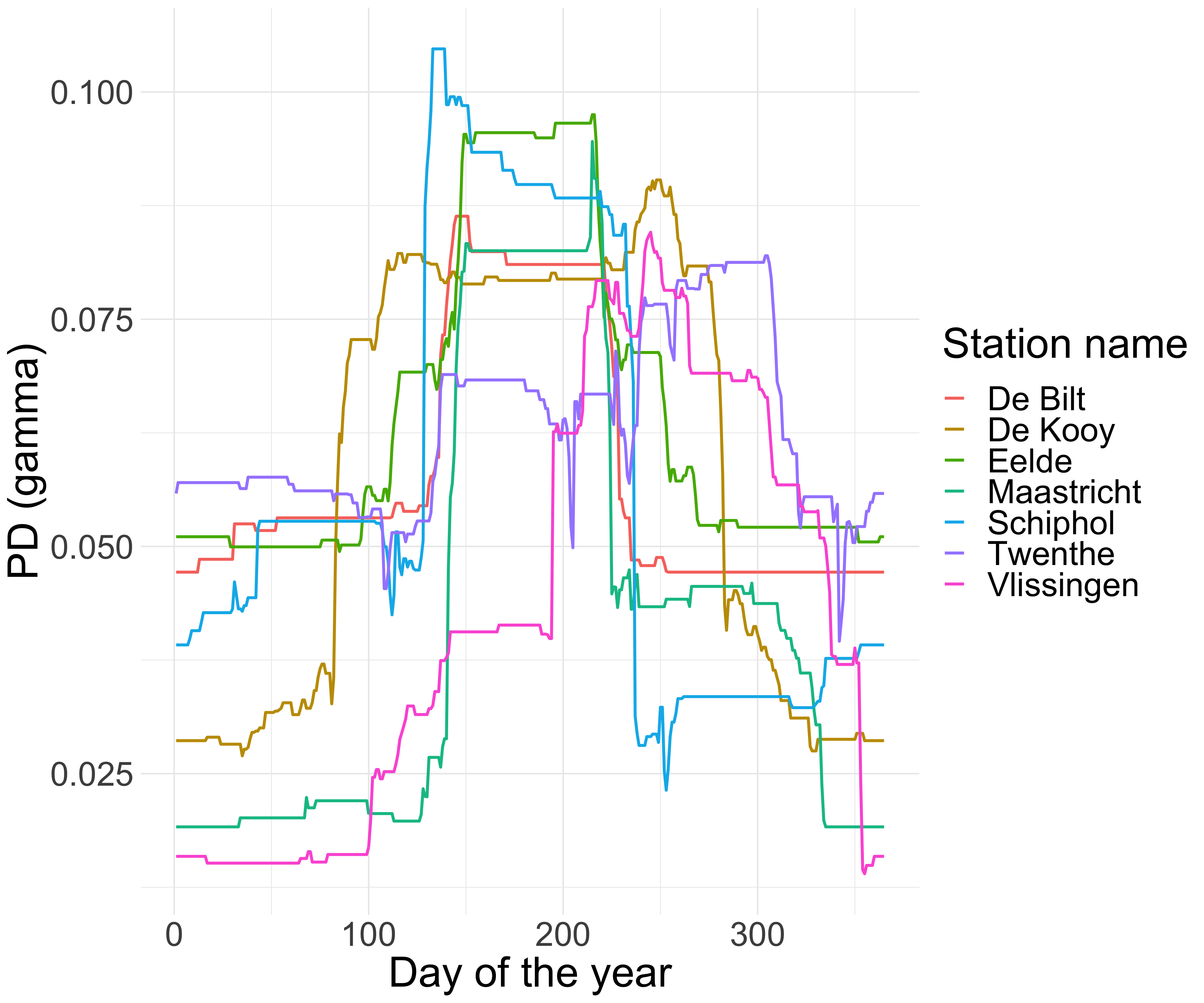}
    \end{minipage}%
    \begin{minipage}{0.5\textwidth}
        \centering
     \includegraphics[width=0.9\linewidth]{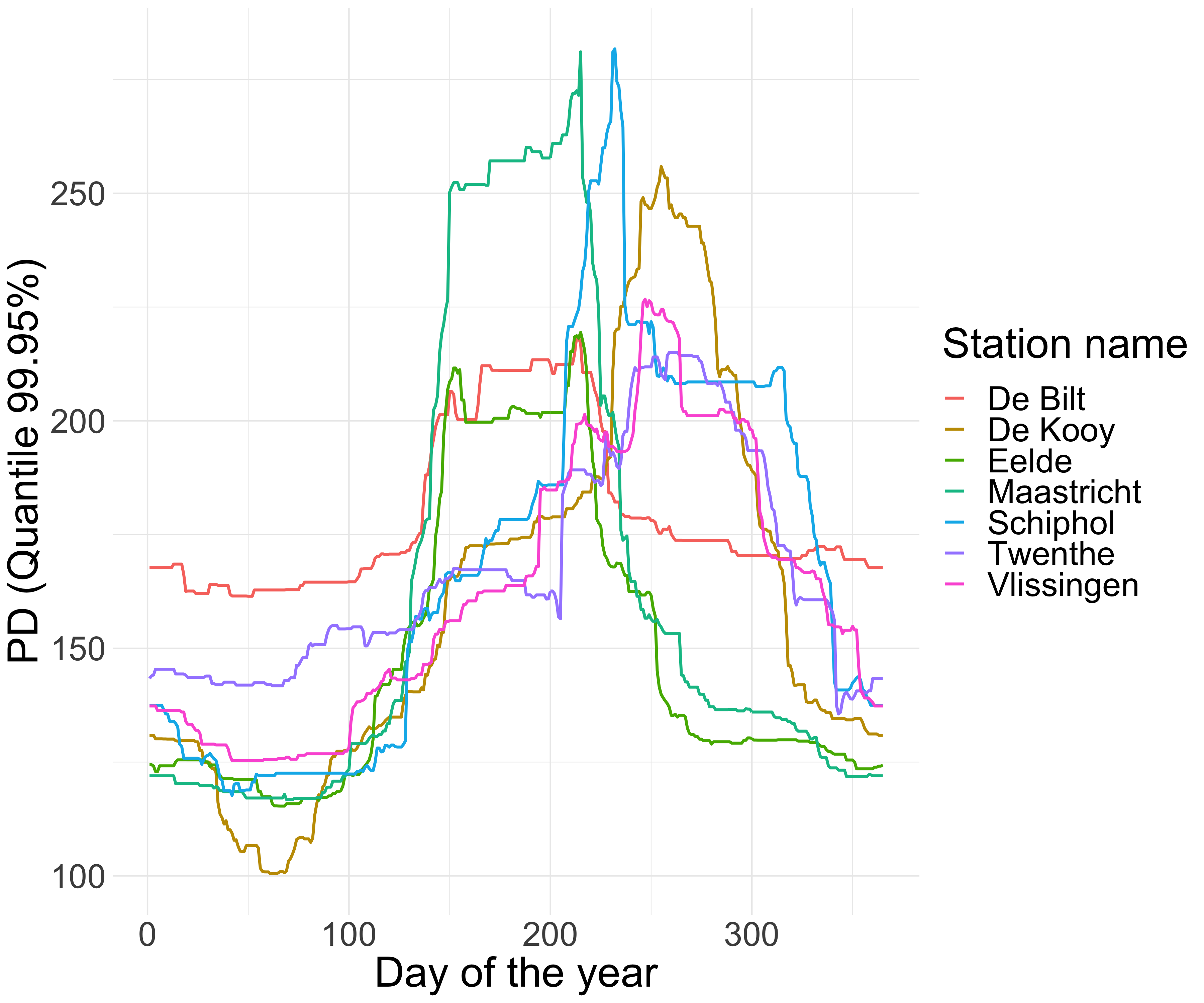}
    \end{minipage}
       \caption{Partial dependence plots of $\hat\gamma^{(l)}$ (left panel) and $\hat Q^{(l)}_{\mathbf{X}}(0.9995)$ (right panel, in 0.1mm) with respect to \texttt{day of year}.}
    \label{fig:PD_doy}
\end{figure}

We first look into the variable importance scores for the fitted models and focus on the relative importance to understand which variables affect the scale and shape parameters, respectively. 
Figure~\ref{fig:VI_relative} shows the relative importance for $\gamma$ and $\sigma$, \tcb{where the scores for the variable \texttt{ens\_sd} and \texttt{ens\_up} correspond to the aggregation of 7 scores (one for each station)}. It is interesting to note that for the shape $\gamma$, the \texttt{day of year} is the most important variable in six out of seven models. This motivates to investigate the seasonality pattern in  the extreme precipitation.  The partial dependence plots of $\hat\gamma^{(l)}$ (left panel) and $\hat Q^{(l)}_{\mathbf{X}}(0.995)$ (right panel) with respect to the \texttt{day of year} are presented in Figure~\ref{fig:PD_doy} for all stations. They indicate that the tail of the precipitation is heavier in summer and autumn than in winter and spring. The curves in the left panel resemble step functions and higher values of $\hat\gamma$ correspond to June, July and August for five stations. For the other two stations Twente and Vlissingen, it is shifted towards autumn.

\begin{figure}%[H]
    \centering
        \centering
     \includegraphics[width=0.9\linewidth]{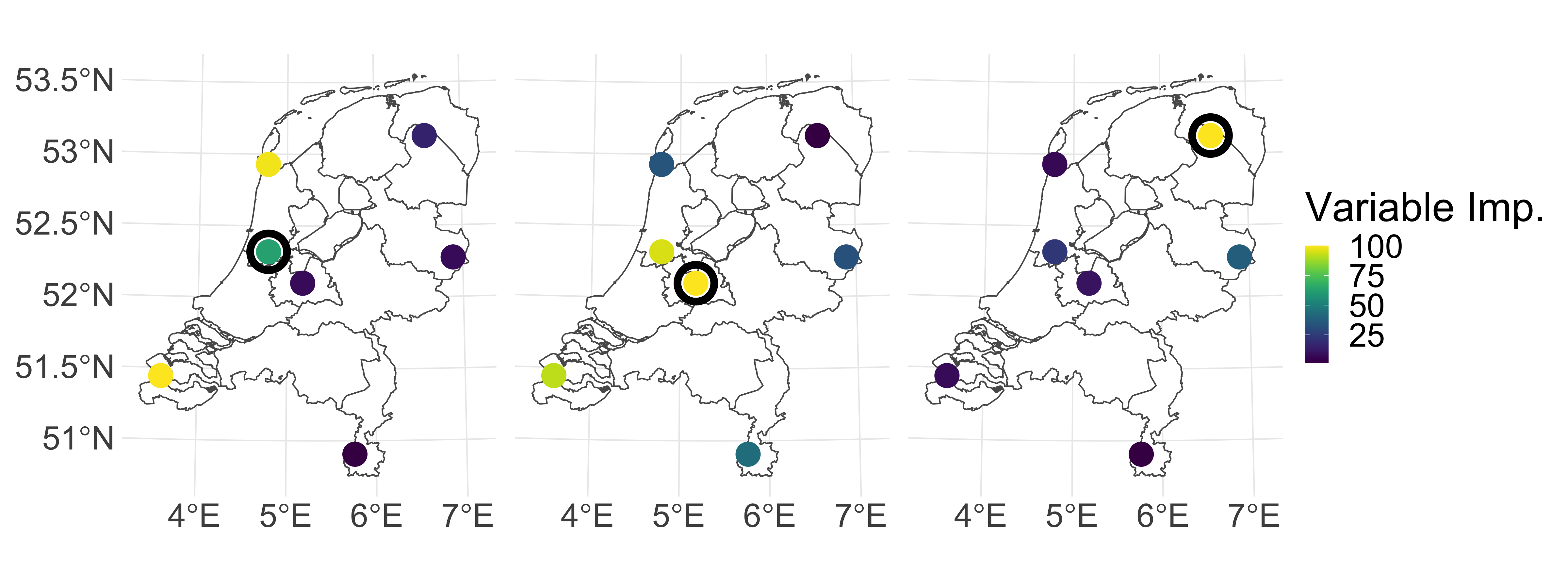}
    \caption{Normalized permutation scores of ensemble statistics per location for three models: Schiphol (left), De Bilt (middle), Eelde (right). 
    The black circle indicates the station for which the model is fitted. From North to South, the stations are: Eelde, De Kooy, Twente, Schiphol, De Bilt, Vlissingen, Maastricht.}
    \label{fig:VI_maps}
\end{figure}
 
Another relevant question concerns the contribution of ensemble statistics of other stations in forecasting the extreme precipitation of a specific location. 
To this end,  we add the permutation scores of ensemble standard deviation and ensemble upper order statistics per station, resulting in seven scores for each model. We then normalize these scores such that the maximum score is 100. The results for three stations are visualized in Figure~\ref{fig:VI_maps}. 
First, quite surprisingly,  when forecasting the extreme precipitation at Schiphol (left plot), the ensemble forecast relies on the information from Vlissingen and De Kooy even more than the information at Schiphol, which might be explained by a coastal effect. Similarly, the model at De Bilt (middle plot) uses the information from Schiphol and Vlissingen. For other stations like Eelde (right plot), the own information of the station is the most important. The maps of the other four stations (De Kooy,  Maastricht, Vlissingen and Twente) are very similar to that of Eelde.

\begin{figure}
\centering
   \includegraphics[width=0.8\linewidth]{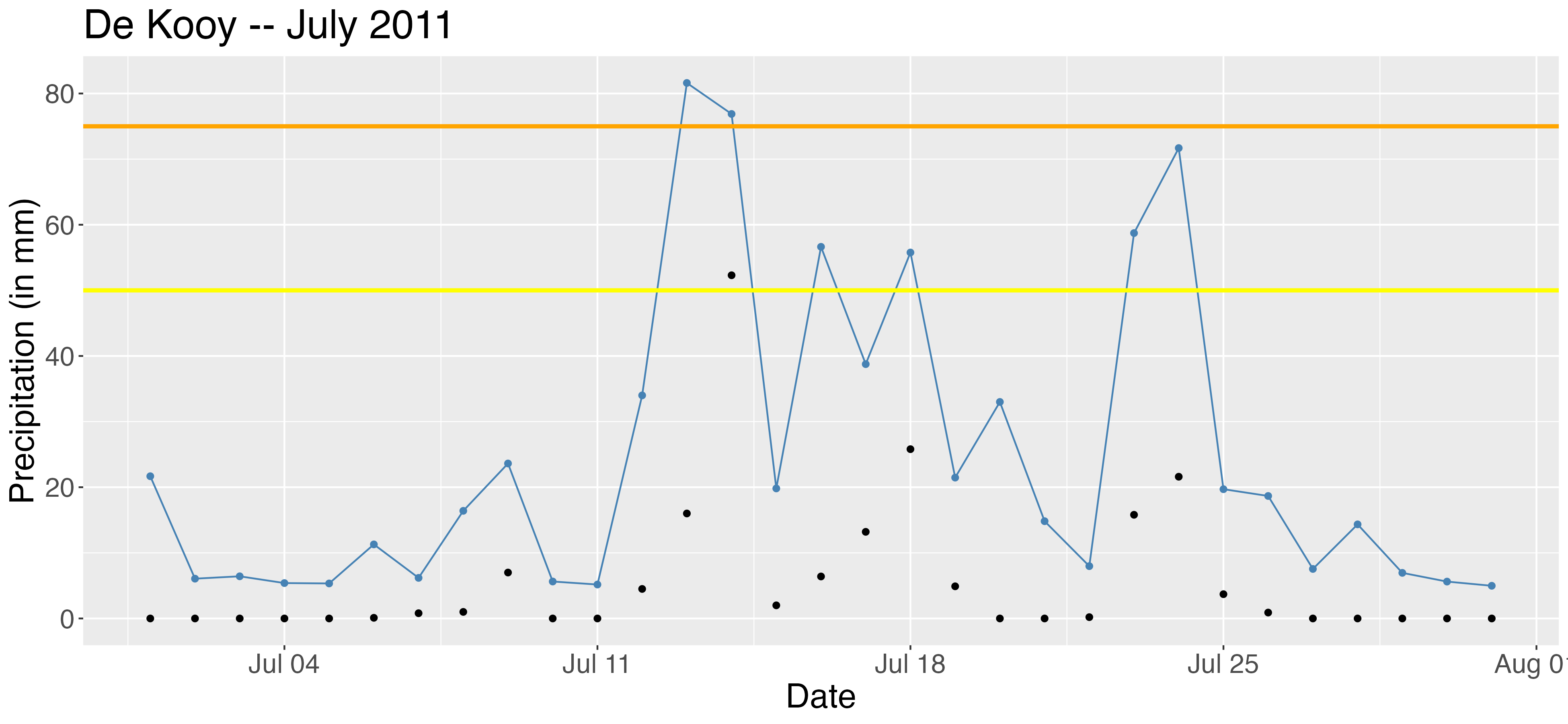}\\
   \includegraphics[width=0.8\linewidth]{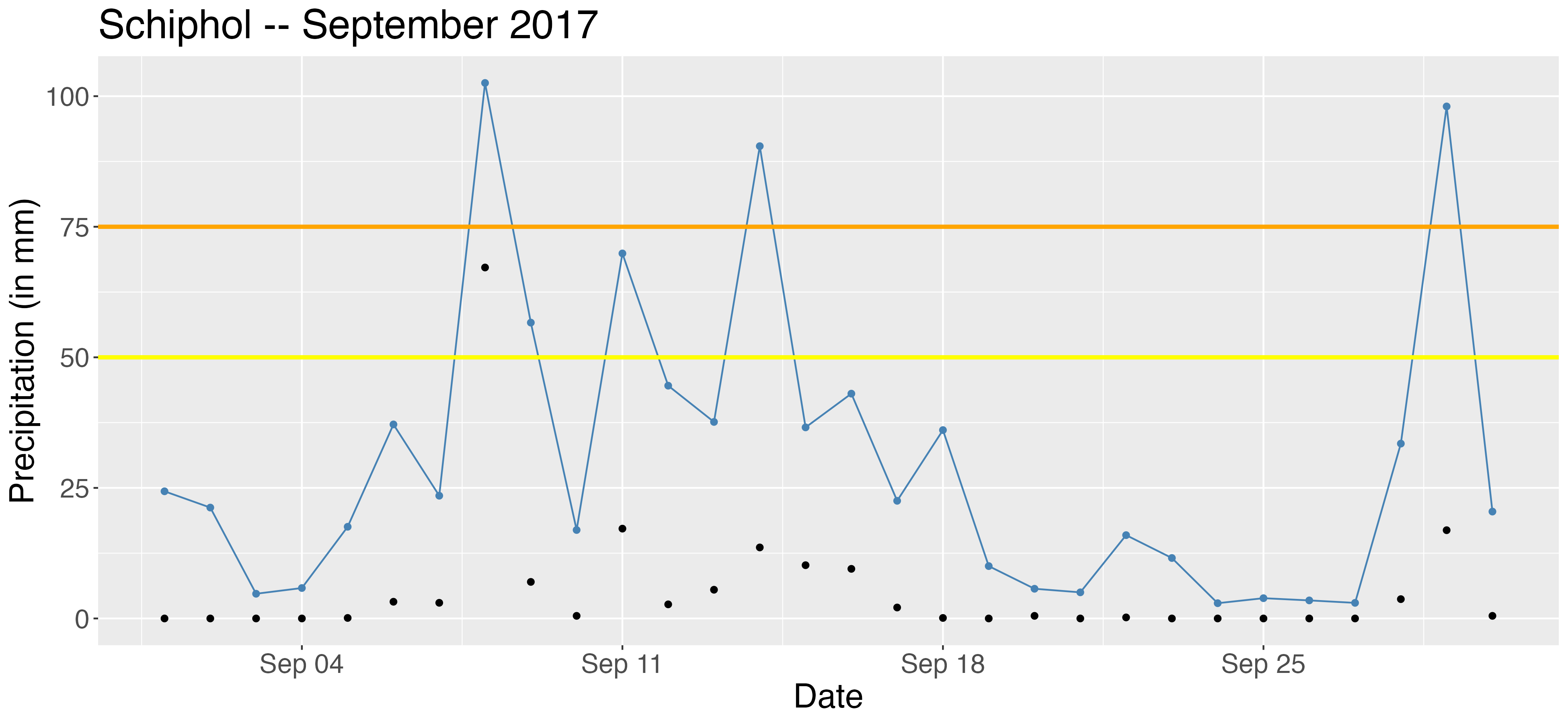}\\
   \includegraphics[width=0.8\linewidth]{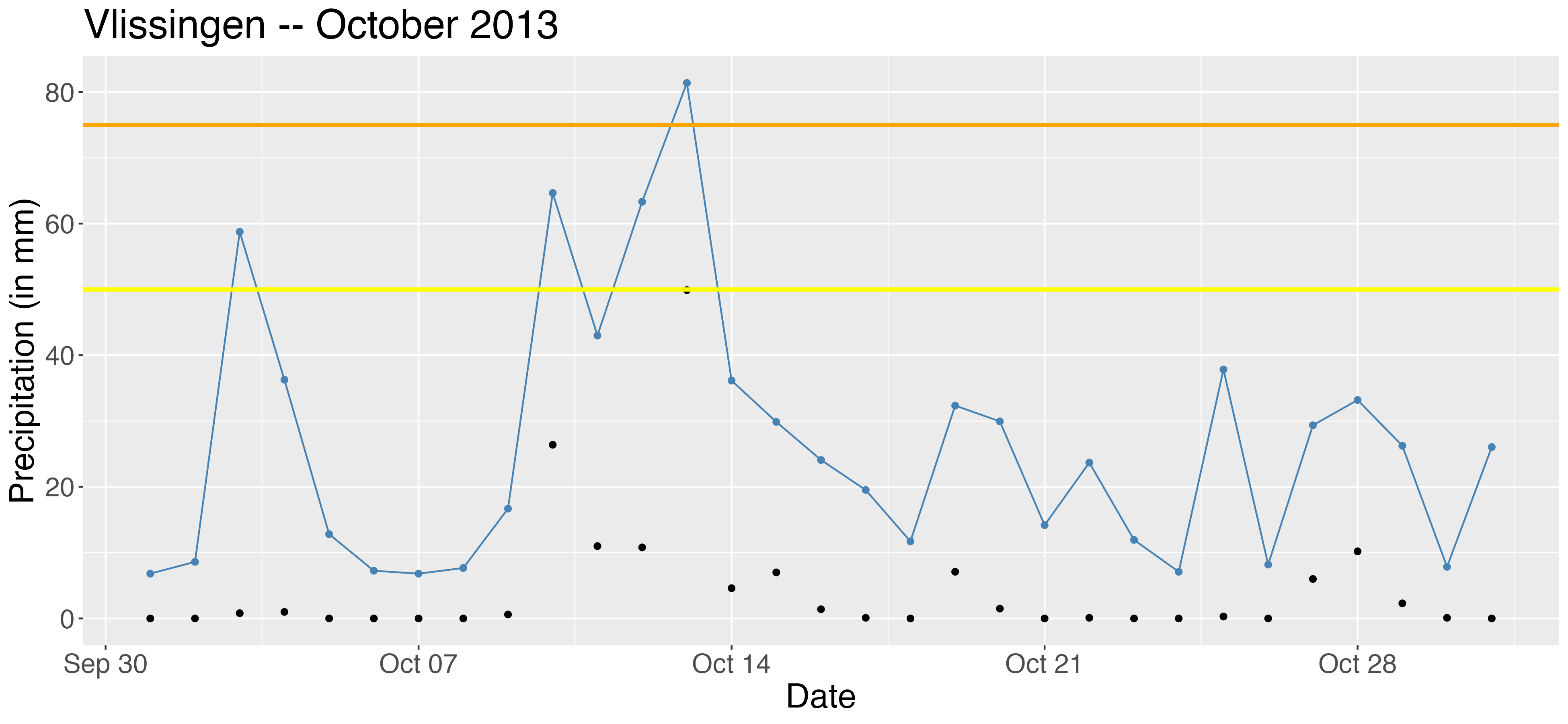}   
\caption{Black points: observed precipitation; blue points: predicted 99.95\% quantile; yellow (orange) line: precipitation threshold of 50 mm (75 mm) for code yellow (orange) weather warnings. }
\label{fig: weather_warning}
\end{figure}

\tcb{
Our method can be used to provide relevant information for weather warning systems.  The Dutch meteorology institute (KNMI) issues three levels of weather warnings  for disruptive weather conditions, namely code yellow, code orange and code red.  Code red, the most severe one, is issued depending on the social impact and safety risk of extreme weather conditions. Code yellow and code orange are issued if some weather quantity such as snowfall, slipperiness, temperature, or wind speed, reaches a specific level. For precipitation, the threshold is 50 mm  (resp.~75 mm) within 24 hours for code yellow (resp. orange). As an illustration on how our method can be informative for the weather warning system, we look into  the predicted 99.95\% quantile by \texttt{gbex} for the month when the  maximum observed precipitation  (over the time span of our data set) occurs, and compare it to the thresholds of code yellow and code orange. Figure \ref{fig: weather_warning} presents the results for three stations: De Kooy, Schiphol and Vlissingen. The maximum observed precipitation were 52.3 mm on July 14, 2011, 67.2 mm on September 8, 2017 and 49.9 mm on October 13, 2013, respectively, for these three stations. For these three days, our prediction of the 99.95\% quantile (using only information form the past) indicates a high level of precipitation, comparable to the code orange level. It could therefore be used for effective early warning. Overall, the blue curve (predicted 99.95\% quantile) is above the black points (observations) and it captures well the days with heavy precipitation.}

\begin{figure}[hbt!]
    \centering
    \begin{minipage}{.5\textwidth}
        \centering
        \includegraphics[width=0.95\linewidth]{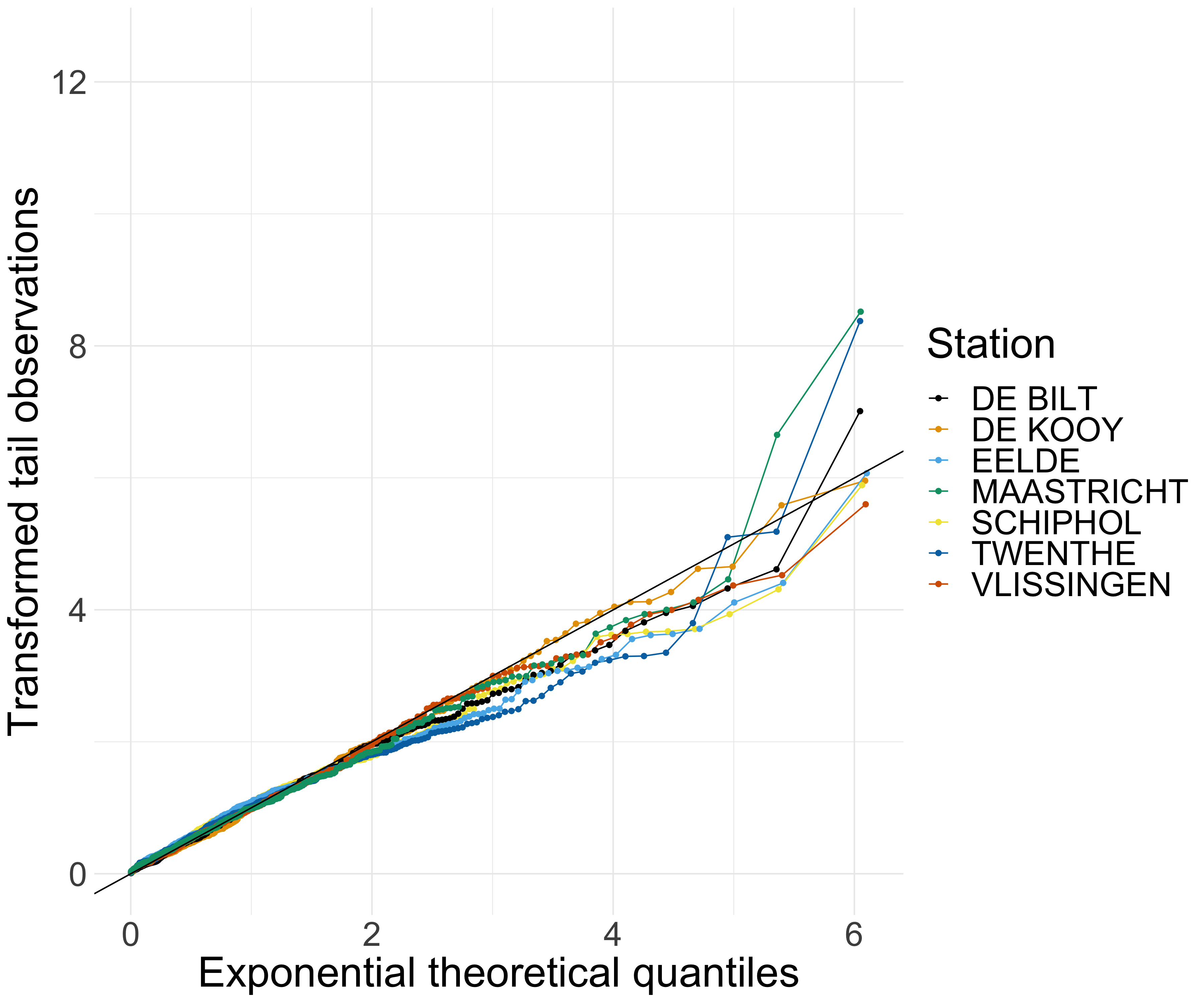}
  
    \end{minipage}%
    \begin{minipage}{0.5\textwidth}
        \centering
     \includegraphics[width=0.95\linewidth]{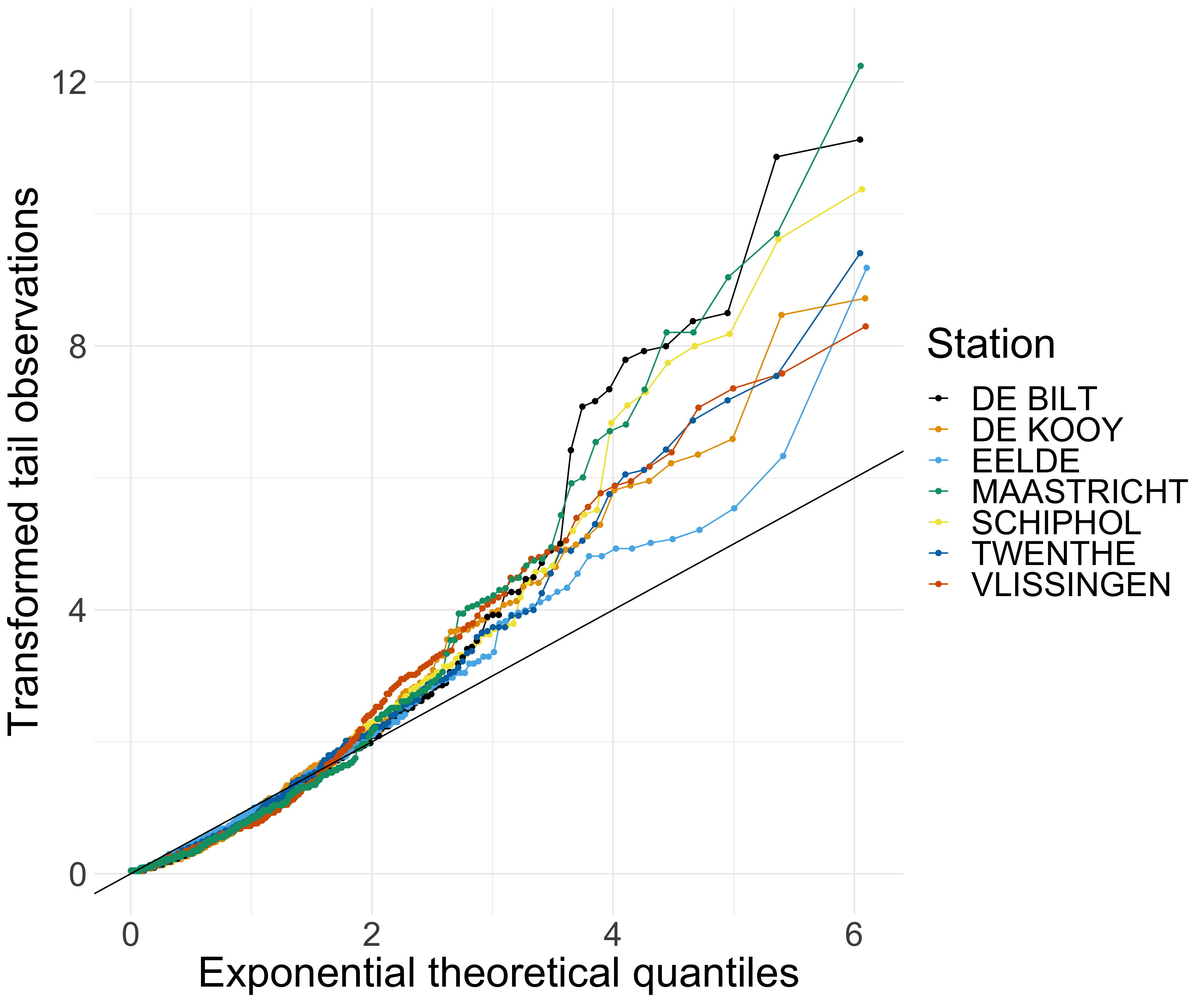}
    \end{minipage}    
    \caption{QQ-plots based on \eqref{eq:qq} for the estimated models at seven stations via  \texttt{gbex} (left panel)  and via the \texttt{constant} method (right panel).}   
    \label{fig:qqplot}
  \end{figure}

We finally assess the goodness of fit of our GPD model and produce QQ-plots comparing the empirical and theoretical quantiles of exceedances above threshold.  We use a transformation to the exponential distribution to compare observations stemming from different stations with different covariate values. More precisely, denoting by $Z_i^{(l)}$  the $i$th exceedance above threshold at station~$l$, then if our model is well-specified
$
Z^{(l)}_i \sim {\rm GPD}\left( \hat \sigma^{(l)}(\bb X_i), \hat\gamma^{(l)}(\bb X_i) \right),
$
and therefore
\begin{equation}
\frac{1}{\hat\gamma^{(l)}(\bb X_i)}\log \left(1+\frac{ \hat\gamma^{(l)}(\bb X_i)Z^{(l)}_i }{\hat \sigma^{(l)}(\bb X_i)} \right)\sim {\rm Exp}(1).  \label{eq:qq}
\end{equation}
The corresponding QQ-plots graphically assess the goodness of fit and we can see in Figure~\ref{fig:qqplot} that the \texttt{gbex} model (left panel) fits the data well at all stations, outperforming the \texttt{constant} model (right panel). \tcb{Such a plot can be used to compare different choices of the intermediate threshold $\tau_0$. Points close to the diagonal indicate that not only the regression model is good, but also that the approximation of the exceedances by the generalized Pareto distribution is appropriate at this this threshold level.}

  \section{Conclusion}  
  The existing literature on extreme quantile regression is so-far limited to low-dimensional predictor spaces \citep{Daouia2013, GardesStupfler2019, Velthoenetal2019} and simple response surfaces \citep{DS90, WangTsai2009, CDD05, you2019}.  
  Our methodological contribution fills a gap in this area. We have developed \texttt{gbex}, a gradient boosting procedure for extreme quantile regression that combines the flexibility of machine learning methods and the rigorous extrapolation from extreme value theory. Our method can handle non-linear complex problems and high-dimensional feature spaces.
  
  We model the tail of the distribution of the response $Y$ by a generalized Pareto distribution (GPD) whose parameters depend on the covariate $\bb X$. Based on exceedances over a high threshold, gradient boosting produces a tree ensemble estimating these parameters using the deviance as the objective function. Tuning parameters can effectively be chosen through our proposed cross-validation, or be fixed to sensible default values. In several numerical experiments we highlight the robustness of \texttt{gbex} against the curse of dimensionality and noise variables. Diagnostic tools are available to quantify the impact of the signal variables on the response. Our method outperforms quantile regression methods from machine learning and classical methods based on extreme value theory. The method can be applied to complex real-world data sets and we show its merits for post-processing of extreme precipitation forecasts in the Netherlands.  

 A very natural yet challenging direction for future research is the theoretical analysis of our gradient boosting procedure. A consistency result for large samples is desirable but all existing results in the literature on gradient boosting assume the convexity of the objective function \citep[e.g.,][]{Biau2021}. The GPD deviance used as objective function in our setting is not convex in the shape parameter $\gamma$. A proper theoretical analysis of \texttt{gbex} therefore seems to be very hard and is outside the scope of the present paper.

 \section*{Acknowledgements}
 This work was supported by the Netherlands Organisation for Scientific Research (NWO) under grant number 14612, by the French Agence Nationale de la Recherche under grant number ANR-20-CE40-0025-01, and  by the Swiss National Science Foundation under grant number 186858.

\section*{Conflict of interest}
The authors declare that they have no conflict of interest.

\section*{Data avalaibility statement}
The datasets generated during and/or analysed during the current study are available in the Github repository \url{https://github.com/JVelthoen/gbex/}. The daily precipitation data used in Section~\ref{sec:application} is also publicly available on \url{http://projects.knmi.nl/klimatologie/daggegevens/selectie.cgi}

\appendix

\section{Likelihood derivatives}\label{app:derivatives}
The gradient boosting algorithm for GPD modeling makes use of the first and second order derivatives of the negative log likelihood $\ell_z(\theta)$, $\theta=(\sigma,\gamma)$ and $z>0$. They are respectively given by
\begin{align*}
\frac{\partial\ell_z}{\partial \sigma}(\theta)&=\frac{1}{\sigma}\Big(1-\frac{(1+\gamma)z}{\sigma+\gamma z}\Big),\\
\frac{\partial\ell_z}{\partial \gamma}(\theta)&=-\frac{1}{\gamma^2}\log\Big(1+\gamma \frac{z}{\sigma}\Big)+\frac{(1+1/\gamma)z}{\sigma+\gamma z},
\end{align*}
and
\begin{align*}
\frac{\partial^2\ell_z}{\partial \sigma^2}(\theta)&= \frac{1}{\sigma(\sigma+\gamma z)}\Big(\frac{z}{\sigma}+\frac{z-\sigma}{\sigma+\gamma z}\Big),\\
\frac{\partial^2\ell_z}{\partial \gamma^2}(\theta)&=   \frac{2}{\gamma^3} \log\big(\gamma\frac{z}{\sigma} + 1\Big) - \frac{2z}{\gamma^2(\sigma + \gamma z)} - \frac{(1+1/\gamma) z^2}{(\sigma + \gamma z)^2}.
\end{align*}

\section{Additional simulation study} \label{app: add sim}
\tcb{
The data generating process is similar to Model 1 in Section \ref{sec:num-exp}. The covariate vector $\bb X \in \RR^{40}$ is distributed uniformly on the cube $[-1, 1]^40$. We consider three heavy-tailed distributions, namely Burr, GPD and  Student's $t$, as the conditional distribution of $Y$ given $\bb X$.   For all models, the scale of $Y$ depends on $\bb X $ through a step function 
\begin{equation}
\mathrm{scale}(\bb X)=1+\mathds{1}(X_1>0). \label{rq: scale}
\end{equation}
The conditional distributions are respectively: }
\begin{itemize}
\item {\bf Model~3:} a Student's $t$-distribution with $2$ degrees of freedom and the scale given in \eqref{rq: scale}. 
\item {\bf Model~4:} a GPD in \eqref{eq:gpd} with $\gamma=0.25$ and $\sigma(\bb x)$ given in \eqref{rq: scale}. 
\item {\bf Models~5-6:}  a Burr distribution with a CDF given by 
$$F(y)=1- \left(1+\left(\frac{y}{\mathrm{scale}(\bb x)}\right)^\alpha \right)^\beta.$$
We choose $\alpha=\beta=2$ for Model 5 and $\alpha=2$, $\beta=1$ for Model 6, which lead to $\gamma=0.25$ and 0.5, respectively. 
\end{itemize}

\begin{figure}[hbt!]
\begin{center}
\includegraphics[width=1\textwidth]{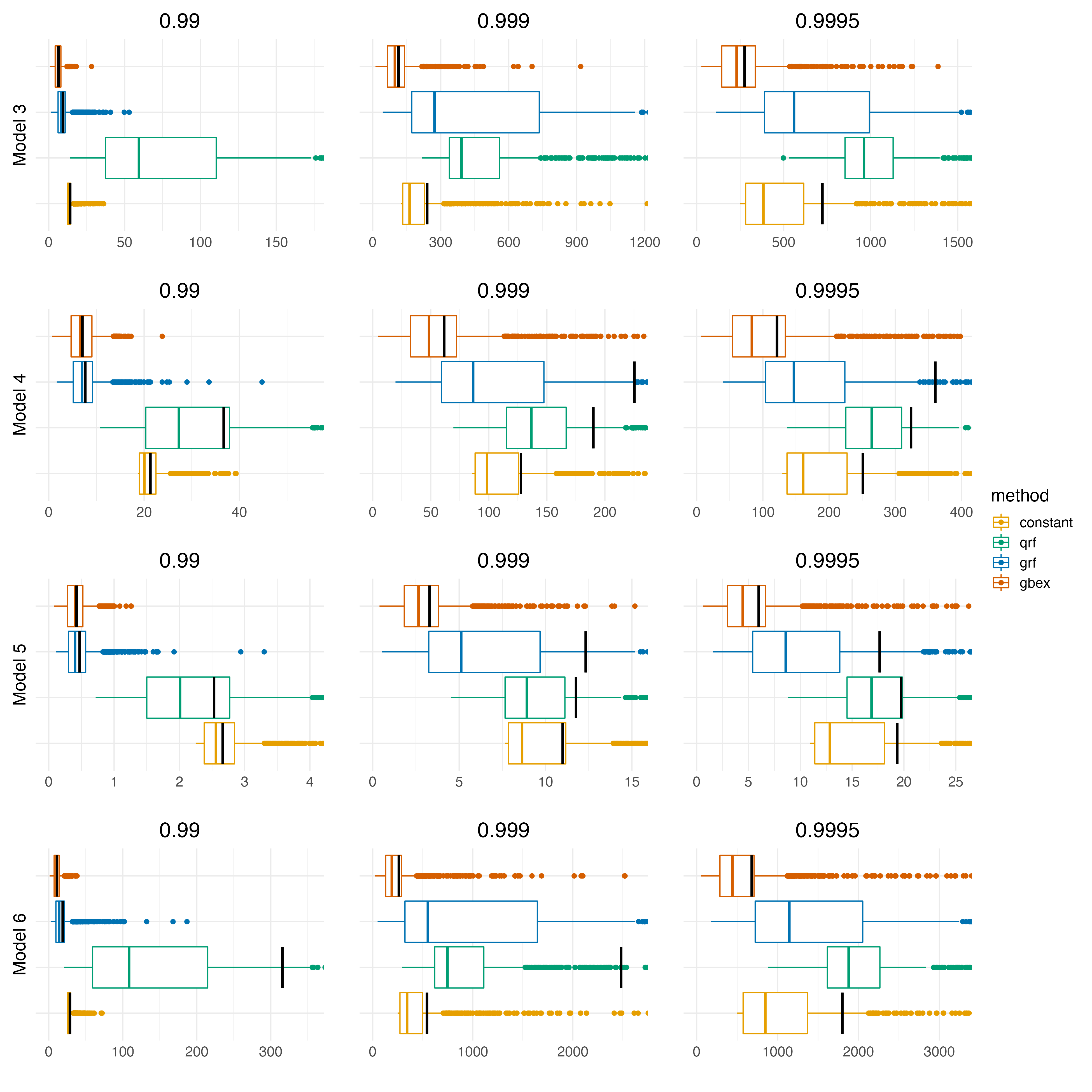}
\caption{Boxplot of ISE based on 1000 replications for the four quantile estimators (\texttt{gbex}, \texttt{grf}, \texttt{qrf} and \texttt{constant}) at different probability levels $\tau=0.99, 0.999, 0.9995$  for Models~3-6.  Some outliers of  \texttt{grf} and  \texttt{qrf} are left out for a clearer comparison.  The black vertical lines indicate the MISE.}
\label{fig: MISE-additional}
\end{center}
\end{figure}

\spacingset{1}

\end{document}